\documentclass[prd,aps,preprint,amsmath,nofootinbib,amssymb,eqsecnum,showkeys,tightenlines]{revtex4}
\pdfoutput=1
\usepackage{verbatim,graphics,graphicx,color,slashed}
\usepackage{ulem} 
\usepackage{hyperref}
\usepackage{soul}

\begin{document}

\title{Indirect and direct signatures of Higgs portal decaying  \\ vector dark matter 
for  positron excess in cosmic rays}

\author{
 Seungwon Baek, 
 P. Ko,
 Wan-Il Park,
 Yong Tang
 }
\affiliation{School of Physics, Korea Institute for Advanced Study,\\
 Seoul 130-722, Korea }
\date{\today}

\begin{abstract}
We investigate the indirect signatures  of the  Higgs portal $U(1)_X$ vector dark matter 
(VDM) $X_\mu$ from both its pair annihilation and decay. 
The VDM is stable at renormalizable level by $Z_2$ symmetry, and  thermalized by 
Higgs-portal interactions.  It can also decay by some nonrenormalizable operators with 
very long lifetime at cosmological time scale.
If dim-6 operators for VDM decays are suppressed by $10^{16}$ GeV scale, the lifetime 
of VDM with mass $\sim$ 2 TeV  is just right for explaining the positron excess 
in cosmic ray observed  by PAMELA and AMS02 Collaborations.  
The VDM decaying into $\mu^+ \mu^-$ can fit the data, evading various constraints 
on cosmic rays. We give one UV-complete model  as an example. 
This scenario for Higgs portal decaying VDM with mass around $\sim2$ TeV can be 
tested by DM direct search at XENON1T, and also at the future colliders by measuring  
the Higgs self-couplings.
\end{abstract}

\pacs{PACS numbers: }

\maketitle

\section{Introdution}

There are convincing evidences of nonbaryonic dark matter (DM) in the universe from 
astrophysical to cosmological scales.
According to the results from Planck~\cite{Ade:2013zuv},  
the dark matter relic density is $\Omega h^2=0.1199\pm 0.0027$ 
with a high precision,  while the standard model (SM) for particle physics has 
no candidate for DM that can account for this measured relic density. 
We need new physics beyond SM (BSM) for (at least) one new particle playing 
the role of nonbaryonic dark matter of the universe. 

Nonbaryonic dark matter must be stable on cosmological time scale. 
In case of decay, its lifetime must be much longer than the age of the Universe.  
The stability of DM is usually 
guaranteed by imposing a discrete global symmetry, such as $Z_2$. 
There are however some arguments that global symmetries may be generically broken 
by quantum gravity\cite{Kallosh:1995hi,Banks:2010zn}, in which case DM 
with global charges would be unstable and decay.   
It can be shown then that the lifetime of DM would be much shorter than the age of the 
Universe using a naive dimensional analysis, if the DM mass is around electroweak 
scale,  $\sim O(100)$ GeV -- O(1) TeV (see Ref.~\cite{Baek:2013qwa} for example).  

Contrary to the global symmetry, a local dark gauge symmetry can be used to guarantee the stability or the longevity of EW scale dark matter.
The simplest model would be adding an extra $U(1)_X$ or some non-Abelian dark 
gauge symmetry to the SM gauge group $G_\mathrm{SM}$ 
(see Ref.s~\cite{Baek:2013qwa,dark_gauge_sym}, for example).  
It is also possible that a hidden sector vector boson could be absolutely 
stable or its lifetime could be much longer than the age of the universe.  
Depending on the structure of a given model, the DM could be scalar, fermion or 
vector particles. 

One interesting scenario is the so-called Higgs portal Abelian vector dark 
matter (VDM) model based on $U(1)_X$ dark gauge symmetry 
(see Ref.~\cite{Djouadi:2011aa}  for example), with an {\it ad hoc} $Z_2$ symmetry 
($X_\mu \rightarrow - X_\mu$) that stabilizes the VDM 
\footnote{Extension with non-abelian dark gauge symmetry is also possible to 
stabilize the VDM~
\cite{Hambye:2008bq,Chen:2009ab,Zhang:2009dd,DiazCruz:2010dc,carone,
Baek:2013dwa}, and fermion DM~\cite{Hur:2007uz,proceeding,Hur:2011sv}. }.
In Ref.~\cite{ko_vdm}, the authors emphasized that it is important to have a 
built-in mechanism for generating the VDM mass by introducing a dark Higgs 
field $\Phi$.  
The new scalar $\Phi$ would interact with the SM particles due to its mixing 
with SM Higgs boson through  the Higgs portal interaction. 
There will be two neutral scalar bosons, the mixtures of the SM Higgs boson and 
the dark Higgs boson.  Due to the generic destructive interference between the 
contributions from two scalar bosons in the amplitude for direct detection 
cross section, constraints from direct detection experiments such as  
XENON100, CDMS and LUX  can be relaxed significantly and the allowed model  parameter space becomes larger than that in the effective model for the Higgs portal VDM~\cite{Djouadi:2011aa}.  
Having a dark Higgs $\Phi$ for the VDM mass, one obtains completely different 
results compared with the effective VDM model where the VDM mass is given by hand 
or by St\"{u}ckelberg mechanism~\cite{Djouadi:2011aa}. 

\begin{figure}
\includegraphics[scale=0.5]{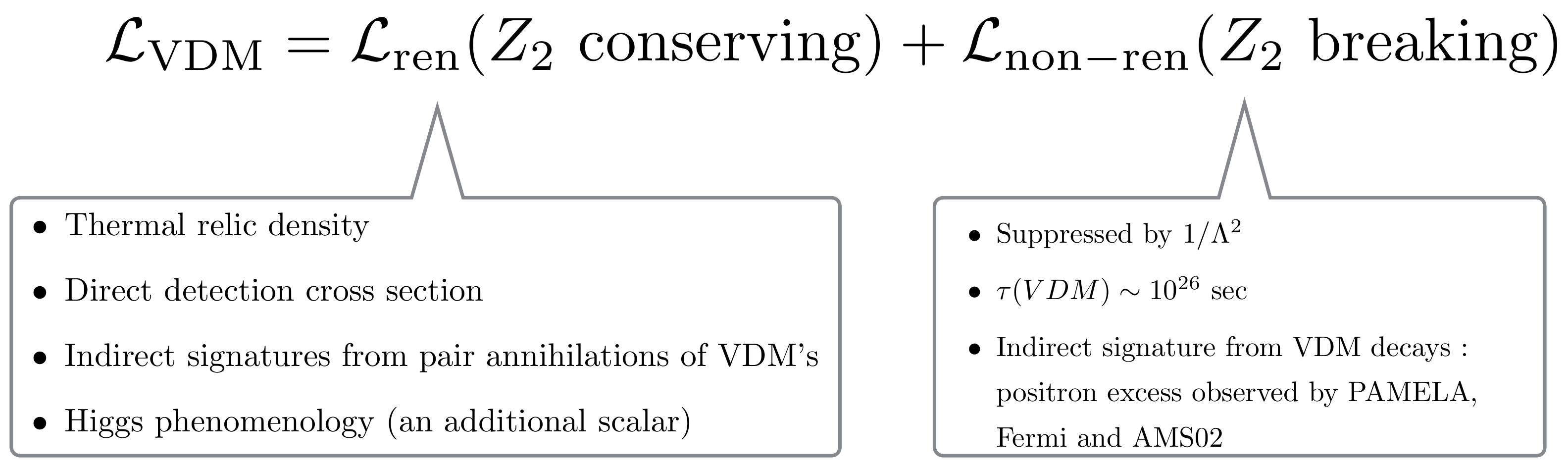}
\caption{Schematic view of the model Lagrangian in this work : the total Lagrangian 
for the VDM is a sum of the  $Z_2$ symmetric renormalizable part and the $Z_2$ 
breaking nonrenormalizable dim-6 operators, neglecting higher dimensional operators. }
\end{figure}

However, the renormalizable VDM model of Ref.~\cite{ko_vdm} may not be 
the complete theory up to Planck scale, although the model was shown to be 
perturbative and the electroweak vacuum could be stable up to Planck 
scale~\cite{ko_vdm}. 
At some scales, $M_{GUT}\simeq 10^{16}$GeV for instance, there could be some new 
physics which can induce higher dimensional operators for the low energy theory 
with $G_\mathrm{SM}\times U(1)_X$ symmetry, which were
not included in Ref.~\cite{ko_vdm} (see Fig.~1). 
Those nonrenormalizable operators can make the VDM decay after electroweak and dark
gauge symmetry breaking, with a resulting DM lifetime that is just at the right order 
in order to explain the recent observed positron excess~\cite{pamela,fermilat,ams02}.

In Ref.~\cite{ko_vdm}, a number of aspects of the renormalizable $U(1)_X$ VDM model 
have been studied in detail, except for the indirect detection signals.  
The purpose of this work is twofold. First of all, we  work out in detail various 
indirect signatures and compare with the cosmic ray data, such as $e^+$, $\bar{p}$, 
$\gamma$ or $\nu$ fluxes. There are two different sources of cosmic rays from the VDM.  
One is the pair annihilations of VDM into the SM particles which are described by 
$Z_2$ symmetric renormalizable Lagrangian of the VDM model constructed in 
Ref.~\cite{ko_vdm} (${\cal L}_{\rm ren}$ in Fig.~1).  
This part will be constrained by thermal relic density, direct detection
and Higgs phenomenology, as described in Ref.~\cite{ko_vdm}. 
The other origin for cosmic rays is the VDM decay into the SM particles 
\footnote{We assume that the ad hoc $Z_2$ symmetry of the renormalizable Lagrangian 
for the VDM is accidental symmetry which can be broken by higher dimensional gauge 
invariant operators, but not by the dim-4 kinetic mixing operators.}, which are 
described by nonrenormalizable higher dimensional operators that break the ad hoc 
$Z_2$-symmetry (${\cal L}_{\rm non-ren}$ in Fig.~1). 

In particular we are interested in explaining the positron excess observed by PAMELA, 
FERMI and AMS02~\cite{fermilat,pamela,ams02}, assuming it has the DM-related origin       
\footnote{It has to be kept in mind that this excess could be also explained 
by astrophysical processes~\cite{Gaggero:2013rya, Blum:2013zsa, DiMauro:2014iia}.}. 
It turns out that the pair annihilation of (V)DM has difficulties to accommodate the 
positron excess,  because the resulting flux is too small compared with the data without
a large boost factor $\sim 10^3$~\cite{Cirelli:2008pk, Baek:2008nz, Bi:2009uj, 
Chen:2009gz, Pearce:2013ola, DeSimone:2013fia, Yuan:2013eja, Cholis:2013psa, 
Jin:2013nta, Yuan:2013eba, Yin:2013vaa,Liu:2013vha, Dev:2013hka,Masina:2013yea, Bergstrom:2013jra, Ibarra:2013zia}. 
In general one has to introduce a large boost factor $\sim 10^3$, which however is 
strongly constrained by the CMB data~\cite{Padmanabhan:2005es, Galli:2009zc, Slatyer:2009yq, Zavala:2009mi, Madhavacheril:2013cna} and Fermi/LAT gamma ray measurements~\cite{Cirelli:2009dv, Chen:2009uq, Calore:2013yia, Ackermann:2012qk, Ackermann:2012rg, Cirelli:2012ut, Gomez-Vargas:2013bea}.
Therefore we are led to consider decays of VDM induced by higher dimensional 
operators. We write down  the complete list of dim-5 and dim-6 operators that 
cause the VDM decays into various SM particles.  Among them, we select operators 
describing VDM decays into lepton pair $l^+ l^-$, and study the positron spectra 
observed by PAMELA and AMS02.  We also present a simple UV completion of 
the nonrenormalizable operators that could account for the positron data  
reported by PAMELA and AMS02 Collaborations. 
In fact a number of works already showed that PAMELA and AMS02 positron excess 
could be fitted{\color{blue},} using $\sim 2$TeV DM decaying into leptons 
\cite{Feng:2013zca, Ibe:2013nka, Kajiyama:2013dba, Feng:2013vva, Dienes:2013lxa, Geng:2013nda}. However thermal relic density or direct detection cross section of the decaying 
DM for PAMELA and AMS02 were (could) not studied, since these issues are independent 
of physics for DM decays explaining PAMELA and AMS02.

In this work we fill this gap by assuming that the decaying VDM for positron excess
were thermalized by the Higgs portal interaction considered in Ref.~\cite{ko_vdm}. 
If we assume that (i) these positron excess is due to the decaying VDM 
of mass $\sim$ 2 TeV which were thermalized by Higgs portal interaction ~\cite{ko_vdm}, 
and (ii) the EW vacuum is stable up to the scale $\Lambda$ where the operators 
for VDM decays~\cite{ko_vdm}, we find that the most parameter region could be 
probed by the future experiments for direct detection of WIMP's in the mass range 
$\sim 2$ TeV.  Also the Higgs self-couplings are modified at the level probed at 
the future colliders such as the ILC. 
Thus we could make a tight connection between the indirect signature of decaying VDM 
(with mass $\sim 2$ TeV) from positron excess in cosmic rays and direct detection of 
such heavy VDM WIMP, as well as the Higgs signal strength and the Higgs-self couplings. 
These important predictions are newly obtained in this work, compared with other works 
on decaying DM for positron excess reported by PAMELA, Fermi and AMS02. 
This accomplishes the second purpose of the present work.  Although we work out in the 
Higgs portal VDM model in this paper, the same strategies could be applied to other 
types of thermal DMs too.

This paper is organized as follows. In Sec.~\ref{sec:model} we give the detailed 
descriptions of the renormalizable part of the Higgs portal VDM model and then give 
theoretical and phenomenological constraints on the parameters for a TeV VDM ($X_\mu$) in Sec.~\ref{sec:tevDM}.  In Sec. ~\ref{sec:anniDM}, we show some examples for cosmic ray 
spectra including gamma ray and neutrinos from the pair annihilation of VDM. 
In Sec.~\ref{sec:higher}, we list the  higher order nonrenormalizable operators 
relevant for the VDM decays into the SM particles, and present the positron spectra 
from the VDM decays.  We also present one UV-complete model for such a dim-6 operator, 
as an illustration.
Then we show the positron spectra and that $X_\mu \rightarrow \mu^+ \mu^-$ 
could fit the positron spectra for $m_X \sim 2$ TeV, for which we identify the parameter 
ranges and discuss other observable effects in direct detection of DM and Higgs 
properties.   Finally we give a summary. 

\section{The Model}\label{sec:model}

We consider a vector dark matter (VDM), $X_{\mu}$, which is associated
with an Abelian dark gauge symmetry $U(1)_{X}$ implemented with discrete $Z_2$ symmetry 
$X_\mu \rightarrow - X_\mu$. The simplest renormalizable and unitary model would 
be the one with an extra complex scalar $\Phi$, whose vacuum expectation 
value (vev) is responsible for the mass of $X_{\mu}$~\cite{ko_vdm}:
\begin{eqnarray}
{\cal L} & = & -\frac{1}{4}X_{\mu\nu}X^{\mu\nu}+(D_{\mu}\Phi)^{\dagger}(D^{\mu}\Phi)-\lambda_{\Phi}\left(\Phi^{\dagger}\Phi-\frac{v_{\Phi}^{2}}{2}\right)^{2}\nonumber \\
 &  & -\lambda_{H\Phi}\left(H^{\dagger}H-\frac{v_{H}^{2}}{2}\right)\left(\Phi^{\dagger}\Phi-\frac{v_{\Phi}^{2}}{2}\right)-\lambda_{H}\left(H^{\dagger}H-\frac{v_{H}^{2}}{2}\right)^{2}+\mathcal{L}_{\mathrm{SM}}.\label{eq:full_theory}
\end{eqnarray}
Here we neglected the kinetic mixing term $X_{\mu\nu}B^{\mu\nu}$ 
\footnote{The issue of $U(1)_X - U(1)_Y$ kinetic mixing is nicely discussed in Refs.~\cite{Chen:2008qs,Chun:2010ve,Choi:2013qra,Davoudiasl:2013jma}.} 
in order to stabilize the VDM at renormalizable interaction level.
The covariant derivative $D_{\mu}$ on $\Phi$ is defined as
\[
D_{\mu}\Phi=(\partial_{\mu}+ig_{X}Q_{\Phi}X_{\mu})\Phi,
\]
 where $Q_{\Phi}$ is the $U(1)_{X}$ charge of $\Phi$ and it can
be rescaled to $\left|Q_{\Phi}\right|=1$.

Assuming the $U(1)_{X}$-charged complex scalar $\Phi$ breaks $U(1)_{X}$ 
spontaneously with a  nonzero vacuum expectation value (VEV) $v_{\Phi}$,
\[
\Phi (x) =\frac{1}{\sqrt{2}}\left(v_{\Phi}+\varphi (x) \right),
\]
the VDM $X_{\mu}$ gets mass equal to $M_{X}=g_{X}|Q_{\Phi}|v_{\Phi}$,
and the hidden sector Higgs field (or dark Higgs field) $\varphi (x)$
will mix with the SM Higgs field $h(x)$ through the Higgs portal interaction, namely the 
$\lambda_{H\Phi}$ term. 
The mixing matrix $O$ between the two scalar  fields is defined as 
\begin{equation}
\left(\begin{array}{c}
h\\
\varphi
\end{array}\right)=\left(\begin{array}{cc}
c_{\alpha} & s_{\alpha}\\
-s_{\alpha} & c_{\alpha}
\end{array}\right)\left(\begin{array}{c}
H_{1}\\
H_{2}
\end{array}\right) \equiv O \left(\begin{array}{c}
H_{1}\\
H_{2}
\end{array}\right) ,
\end{equation}
 where $s_{\alpha}(c_{\alpha})\equiv\sin\alpha(\cos\alpha)$, $H_{i}(i=1,2)$
are the mass eigenstates with masses $M_{H_i}$. $H_{1}$ will be identifid
as the $125$GeV Higgs boson observed  at the LHC throughout this paper.
The mass matrix of two scalar bosons in the basis $(h,\varphi)$ can be written in terms of 
either Lagrangian parameters or the physical parameters as follows:
\begin{equation}
\mathcal{M}\equiv
\left(\begin{array}{cc}
2\lambda_{H}v_{H}^{2} & \lambda_{H\Phi}v_{H}v_{\Phi}\\
\lambda_{H\Phi}v_{H}v_{\Phi} & 2\lambda_{\Phi}v_{\Phi}^{2}
\end{array}\right)=\left(\begin{array}{cc}
M_{H_1}^{2}c_{\alpha}^{2}+M_{H_2}^{2}s_{\alpha}^{2} & \left(M_{H_2}^{2}-M_{H_1}^{2}\right)s_{\alpha}c_{\alpha}\\
\left(M_{H_2}^{2}-M_{H_1}^{2}\right)s_{\alpha}c_{\alpha} & M_{H_1}^{2}s_{\alpha}^{2}+M_{H_2}^{2}c_{\alpha}^{2}
\end{array}\right).\label{eq:mass_matrix}
\end{equation}
The mixing angle $\alpha$ of two scalar bosons is determined by
\begin{eqnarray*}
\tan2\alpha & = & \frac{2\mathcal{M}_{12}}{\mathcal{M}_{22}-\mathcal{M}_{11}},\;\mathrm{or}\;\sin2\alpha=\frac{2\lambda_{H\Phi}v_{H}v_{\Phi}}{M_{H_2}^{2}-M_{H_1}^{2}}.
\end{eqnarray*}

This renormalizable Lagrangian for the Higgs portal VDM model, Eq.~(2.1), 
has four more parameters compared with the SM: 
$\lambda_\Phi$, $v_{v_\Phi}$, $g_X$ and $\lambda_{H\Phi}$. 
For convenience, we shall trade them with the following set of input parameters:  
$M_X$, $M_{H_2}$,  $g_X$ and $\sin{\alpha}$.  
Since our aim is to explain  the positron excess observed by PAMELA and AMS02 
in terms of VDM decays, we will concentrate mainly on heavy VDM with mass around 
a few TeV in this paper.  

\section{$\mathcal{O}$(TeV) VDM and Phenomenological Constraints}\label{sec:tevDM}

\begin{figure}
\includegraphics[width=0.48\textwidth,height=0.48\textwidth]{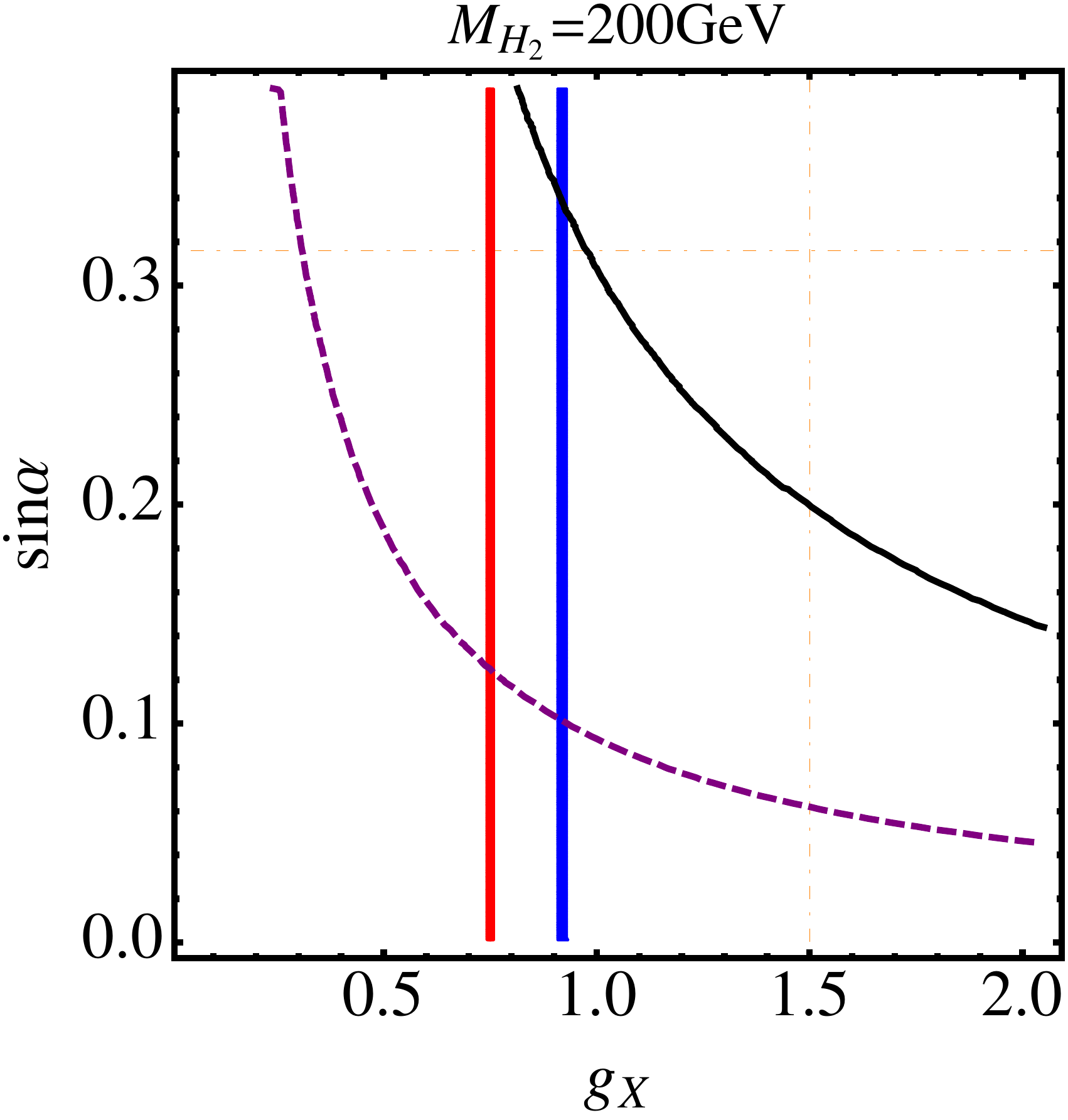}
\includegraphics[width=0.48\textwidth,height=0.48\textwidth]{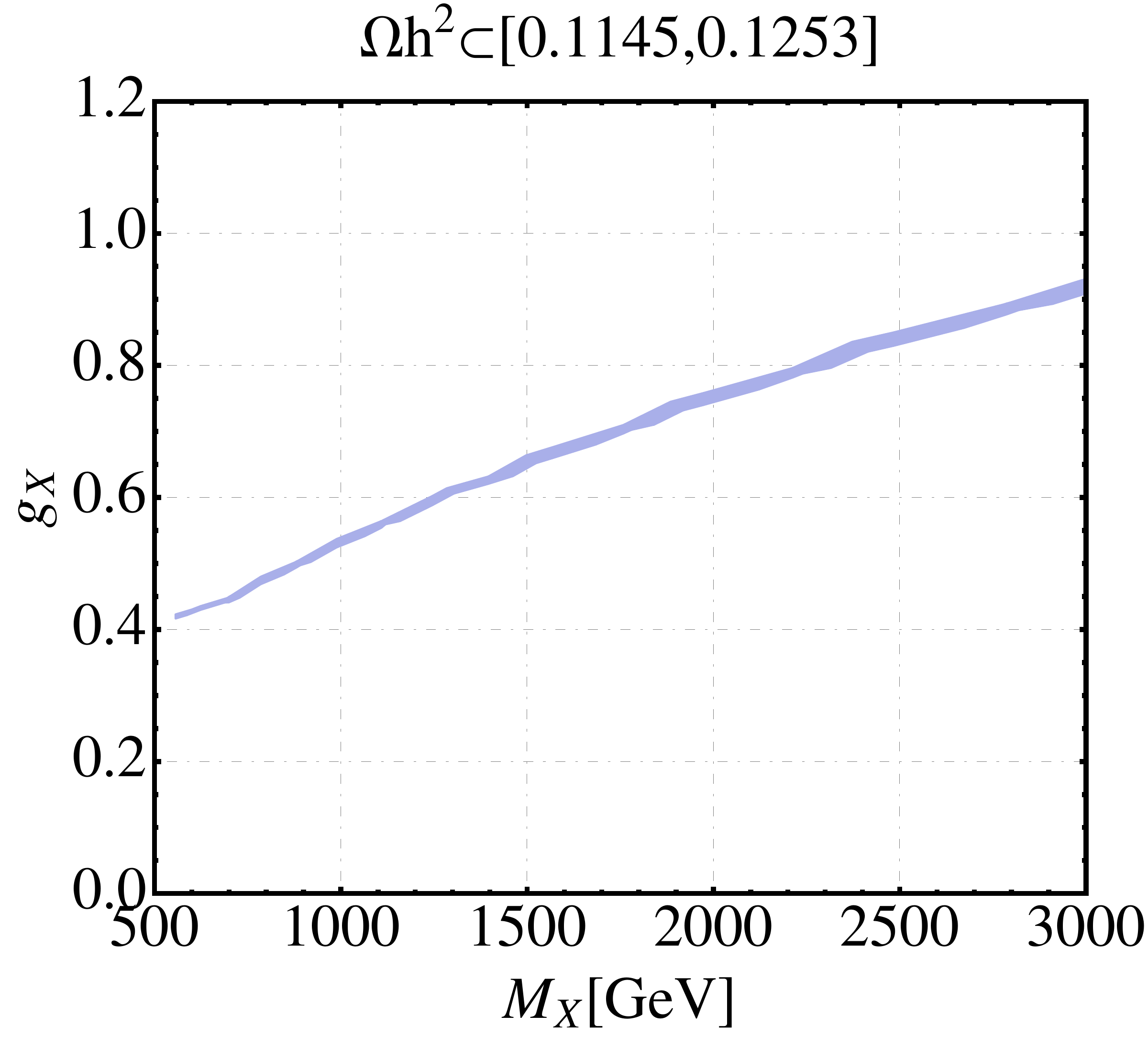}
\caption{(Left panel)The horizontal dot-dashed line set the boundary for $\sin\alpha\simeq 0.32$ or $\sin^{2}\alpha\simeq0.1$ from Higgs data. The vertical dot-dashed one marks the limit for perturbativity.
The solid(dashed) curves corresponds $\sigma_{XN}=10^{-44}\left(10^{-45}\right)$
$\mathrm{cm}^{2}$.
The vertical red and blue bands set the correct relic density for
$M_{X}=2{\rm TeV,\;3{\rm TeV}},$ respectively. (Right panel)This plot shows the relation between $g_X$ and $M_X$ constrainted by $\Omega h^2$, the blue band region is allowed with $2\sigma$ variation. \label{fig:constraints} }
\end{figure}

\begin{figure}[t]
\includegraphics[width=0.48\textwidth,height=0.48\textwidth]{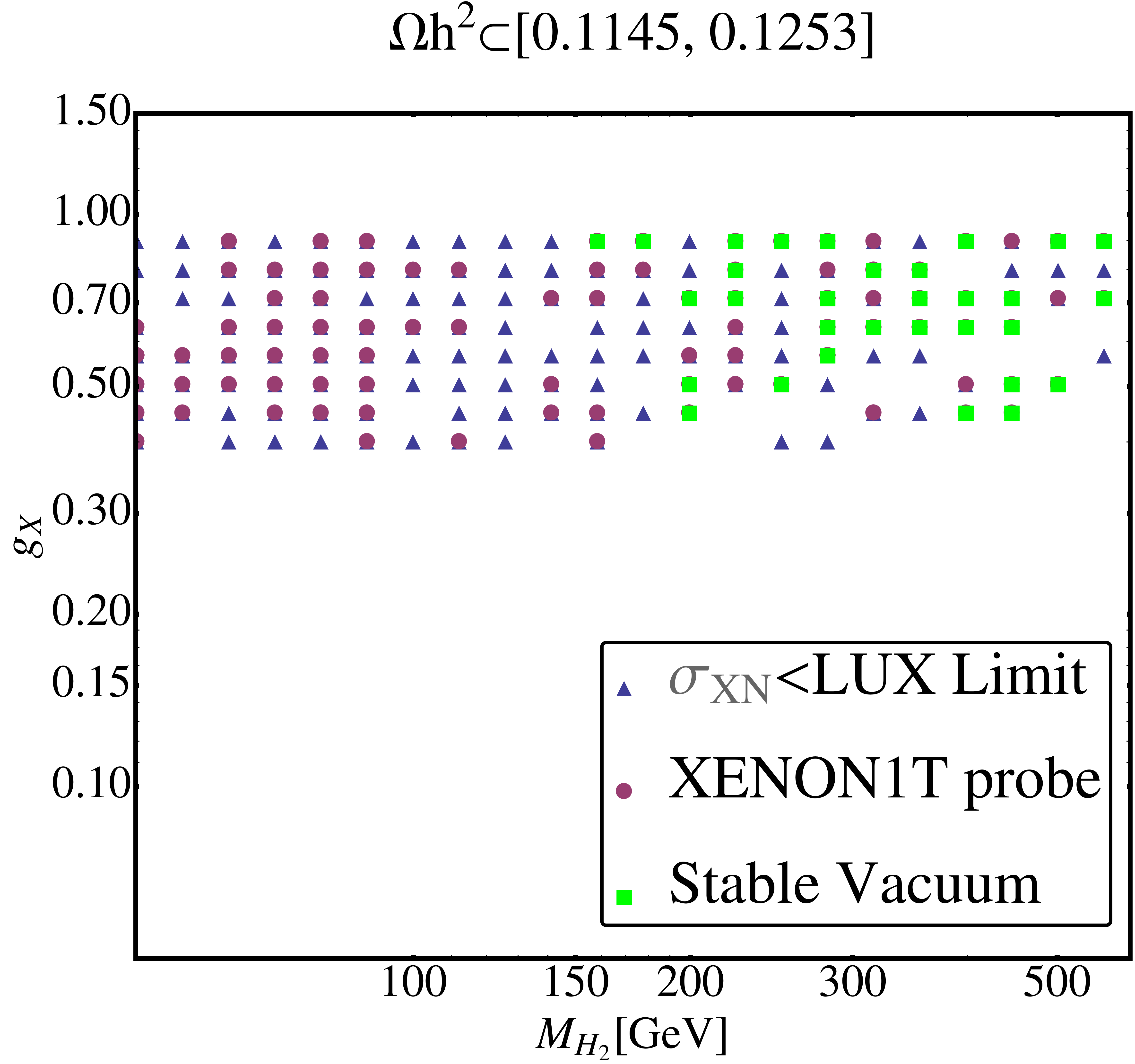}
\includegraphics[width=0.48\textwidth,height=0.475\textwidth]{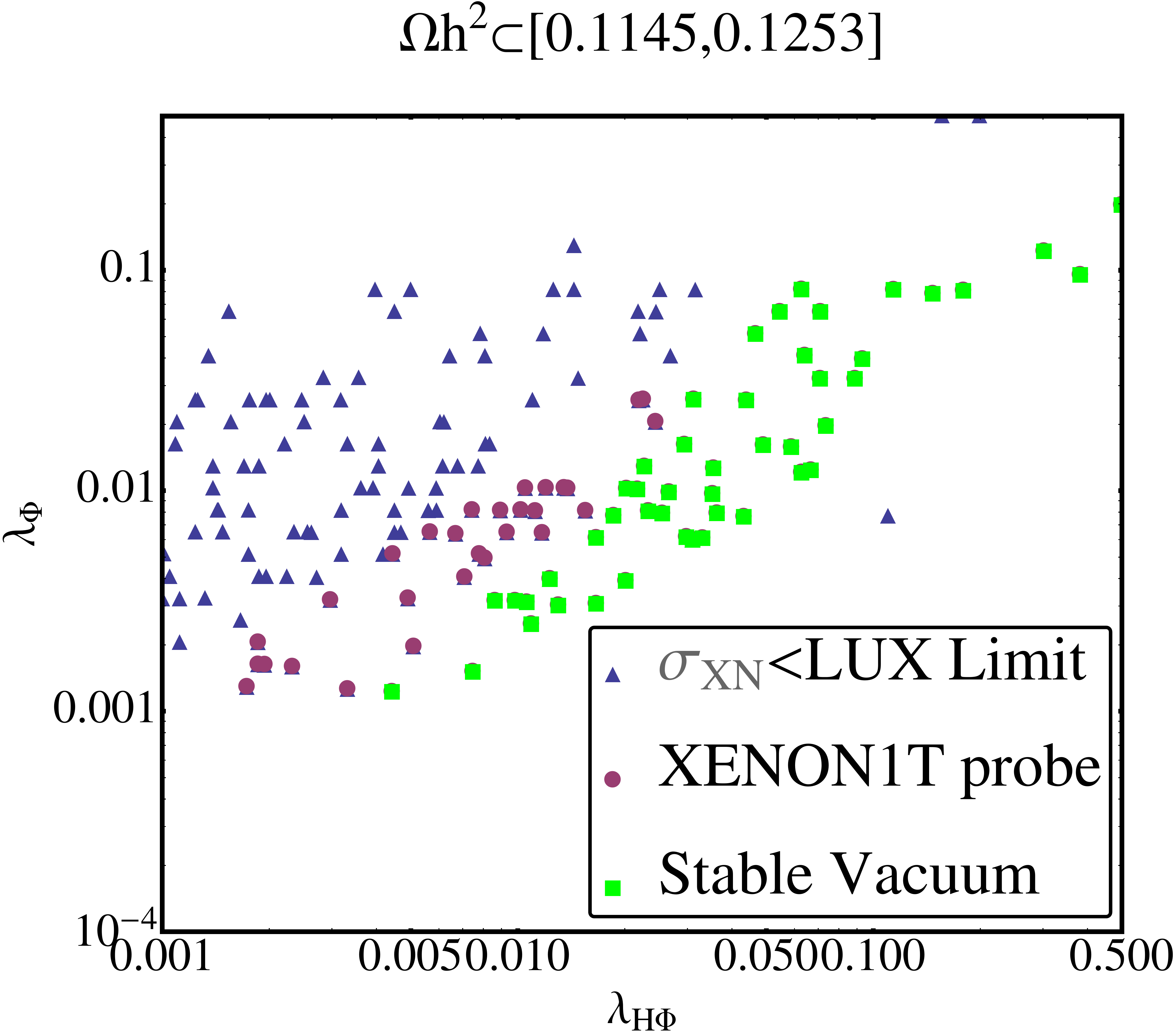}
\caption{Scatter plots with different axis for $0.5 \mathrm{TeV} < M_X < 3 \mathrm{TeV}$. Every point satisfies the relic density in $2\sigma$. Blue triangles are below the LUX Limit and purple circles can be probed by dark matter direct search in the near future. Green squares give stable EW vacuum . It can be seen that all green squares can be probed by the XENON1T.\label{fig:scatter}}
\end{figure}

For a successful explanation of the positron excess reported by PAMELA and AMS02, 
we need a dark matter around $\mathcal{O}$(TeV).   Therefore we first would like to 
show that such a heavy VDM can still be compatible with various constraints from 
colliders, thermal relic density, theoretical consistencies, {\it etc.}. 
Figs.~\ref{fig:constraints},~\ref{fig:scatter} and~\ref{fig:xenon} show that 
there is indeed an ample parameter space for accommodating a TeV VDM.  
In this section, we shall provide detailed discussions on  various relevant constraints 
on the renormalizable model Lagrangian Eq.~(2.1) one by one. The indirect signature 
from the renormalizable model and from higher dimensional operators will be discussed 
in Sec.~IV and Sec.~V, respectively.  

\subsection{Constraint from Higgs data}

The current LHC data on the Higgs signal strengths in various production and decay  
channels give a constraint on the mixing angle, 
$|\sin\alpha|\lesssim 0.32$ or $\sin^{2}\alpha\lesssim 0.1$~\cite{ko_jung}.  
In the Fig.~\ref{fig:constraints}, the region above the horizontal dot-dashed line 
yields $\sin\alpha>0.32$, and therefore is disfavored by the current LHC data.
Then assuming the scalar mixing angle $\alpha$ is small, we can make an approximation: 
\[
\cos\alpha\simeq1  , \  {\rm and} \ \
\sin\alpha\simeq\frac{\lambda_{H\Phi}v_{H}v_{\Phi}}{M_{H_2}^{2}-M_{H_1}^{2}}.
\]  
When $M_{X}\sim2$ TeV, the relic density constrains the gauge coupling to be around 
$g_{X}\sim0.7$, as shown in Fig.~\ref{fig:constraints}. If we further assume that $H_{2}$ 
is still in thermal equilibrium before the VDM  freezes out, the $H_2$ mass should be 
smaller than $M_{X}$. Taking $M_{H_2}\sim500$ GeV and a tiny mixing angle $\alpha$ 
for example, we have 
\begin{eqnarray*}
\lambda_{H\Phi} & \sim & \frac{\sin\alpha\left(500^{2}-125^{2}\right)}{246\times2000}\sim0.5\times\sin\alpha,\;
\nonumber  \\
\lambda_{\Phi} & \sim & \frac{500^{2}}{2\times2000^{2}}=0.03, 
\nonumber   \\
\lambda_{H} & = & \frac{125^{2}+\left(500^{2}-125^{2}\right)\sin^2\alpha}{2\times246^{2}}\simeq0.13+2\sin^2\alpha.\label{eq:smallness}
\end{eqnarray*} 
This gives only a rough estimate of approximate values for the parameters.

Throughout this section, we restrict the parameters in the following ranges: 
\begin{align*}
0.5\textrm{TeV} <  & M_{X}        < 3\textrm{TeV},\;\\
1\textrm{GeV}   <  & M_{H_2}      < 600\textrm{GeV},\;\\
0.4             <  & g_X          < 1.0,\; \\
0.001           <  & \sin^2\alpha < 0.1.
\end{align*}
When scanning over these parameters,  we take flat distributions in $25$ steps for $M_{X},M_{H_2},\sin\alpha$ with logarithmic metric, and $g_X$ with linear metric.
The viable and exact values of $M_{X}$ and $g_{X}$ are further constrained by thermal 
relic density and perturbativity conditions as discussed below. 
Distributions of the viable points are illustrated in Figs.~\ref{fig:constraints},
\ref{fig:scatter},\ref{fig:xenon} and \ref{fig:hself}.

The mixing angle $\alpha$ is also constrained by DM direct search and 
the lifetime of $H_2$ as shown in the right panel of Fig.~\ref{fig:xenon}. 
The upper bounds are from XENON100~\cite{xenon100} (red), LUX~\cite{lux} (orange), 
and vacuum stability (blue) for 2 TeV VDM as examples, where the red and orange 
regions are excluded.  The EW vacuum becomes absolutely stable in the blue region. 
The lower bound on the mixing angle $\alpha$ comes from the BBN constraint 
on the lifetime of $H_2$ , where we require $H_2$'s lifetime $\tau_{H_2}<10^{-2}$s.  Otherwise 
a long-lived $H_2$ could be dangerous to the successful BBN for very small $\sin \alpha$.  
It turns out that thermalization of the dark sector puts a much more stringent lower bound 
than BBN except in the low $M_{H_2}$ region. We shall discuss this case later in detail.

\subsection{Thermal relic density }

\begin{figure}[htb]
\includegraphics[width=1\textwidth]{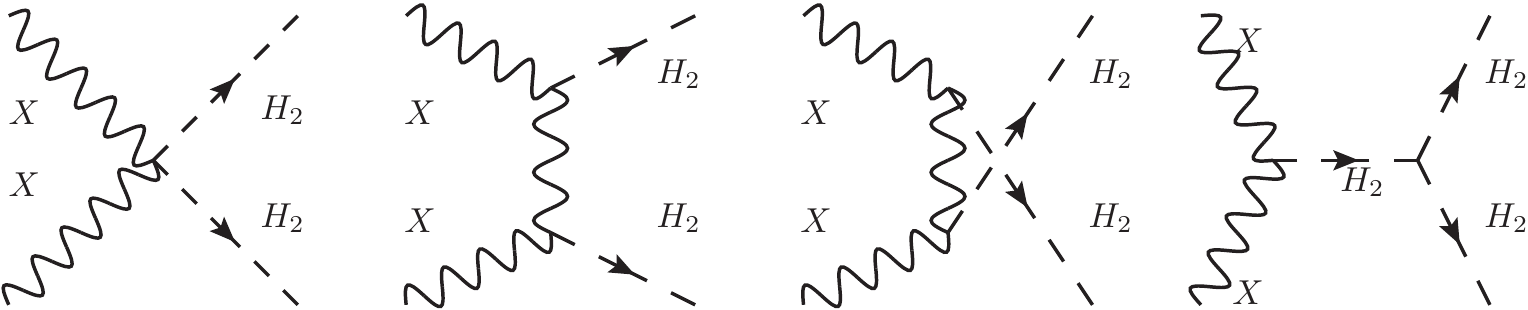}
\caption{Main feynman diagrams for annihilation $XX\rightarrow H_2 H_2$, with vertex functions, $g_{XXH_2}\propto M_X$ and $g_{XXH_2H_2}\propto M^2_X/v^2_{\Phi}$. The last one can be neglected due to the smallness
of $\lambda_{\Phi}.$\label{fig:diagrams} }
\end{figure}
We are interested in the parameter space, $M_{X}\sim O(1)$ TeV and 
$M_{H_{2}}\sim\mathcal{O}(100)$ GeV, aiming at explaining the positron excess. 
As a result,   $\lambda_{\Phi}$ and $\lambda_{H\Phi}$ are small enough that only the first 
three Feynman diagrams of Fig.~\ref{fig:diagrams} need to be considered for $X_{\mu}$ annihilation. For heavy $X_\mu$ and small mixing between $H_1$ and $H_2$, the dominant annihilation channel is $X X \rightarrow H_2 H_2 $.  Then the quantity 
$\sigma v$ relevant to thermal relic density is calculated as
\begin{eqnarray}
\sigma v & = & \frac{1}{3\times3\times2}\frac{1}{2M_{X}\sqrt{s}}\int\frac{\left|\mathcal{M}\right|^{2}}{\left(4\pi\right)^{2}}\frac{\left|p_{1}\right|}{\sqrt{s}}d\Omega\nonumber \\
 & \simeq & \frac{g_{X}^{4}}{144\pi M_{X}^{2}}\left[3-\frac{8\left(M_{H_{2}}^{2}-4M_{X}^{2}\right)}{M_{H_{2}}^{2}-2M_{X}^{2}}+\frac{16\left(M_{H_{2}}^{4}-4M_{H_{2}}^{2}M_{X}^{2}+6M_{X}^{4}\right)}{\left(M_{H_{2}}^{2}-2M_{X}^{2}\right)^{2}}\right],\label{eq:sigmav}
\end{eqnarray}
where $\frac{1}{3\times3\times2}$ accounts for the averaging over
polarizations for initial states and identical factor for final states, $s\simeq 4M^2_X$ at decoupling time,
and 
\[
\left|\mathcal{M}\right|^{2}=g_{X}^{4}\left[12-\frac{32\left(M_{H_{2}}^{2}-4M_{X}^{2}\right)}{M_{H_{2}}^{2}-2M_{X}^{2}}+\frac{64\left(M_{H_{2}}^{4}-4M_{H_{2}}^{2}M_{X}^{2}+6M_{X}^{4}\right)}{\left(M_{H_{2}}^{2}-2M_{X}^{2}\right)^{2}}\right].
\]
Since  $\sigma v$ is independent of $v$ at the leading order in $v$, we can replace the 
thermal averaged $\langle\sigma v\rangle$ with Eq.~(\ref{eq:sigmav}) in the calculation 
of relic density.   For $\langle\sigma v\rangle\sim3\times10^{-26}{\rm cm^{3}s^{-1}}$,
$M_{X}\sim{\rm TeV}$ and $M_{H_{2}}\ll M_{X}$, we have
\begin{equation}
g_{X}\sim0.57\times\left(\frac{M_{X}}{1{\rm TeV}}\right)^{\frac{1}{2}}.\label{eq:gxmx}
\end{equation}
As shown in Fig.~\ref{fig:constraints}, the red and blue vertical
bands display the correct relic density ($\Omega h^2=0.1199\pm 0.0027$~\cite{Ade:2013zuv}) of DM for $M_{X}=2$ TeV  and $M_{X}=3$ TeV, respectively. The precise relation between $g_X$ and $M_X$ is shown in the right panel of Fig.~\ref{fig:constraints}, where we used $\texttt{micrOMEGAs3.1}$~\cite{micromegas} for the numerical calculation.

\begin{figure}
\includegraphics[width=0.48\textwidth,height=0.462\textwidth]{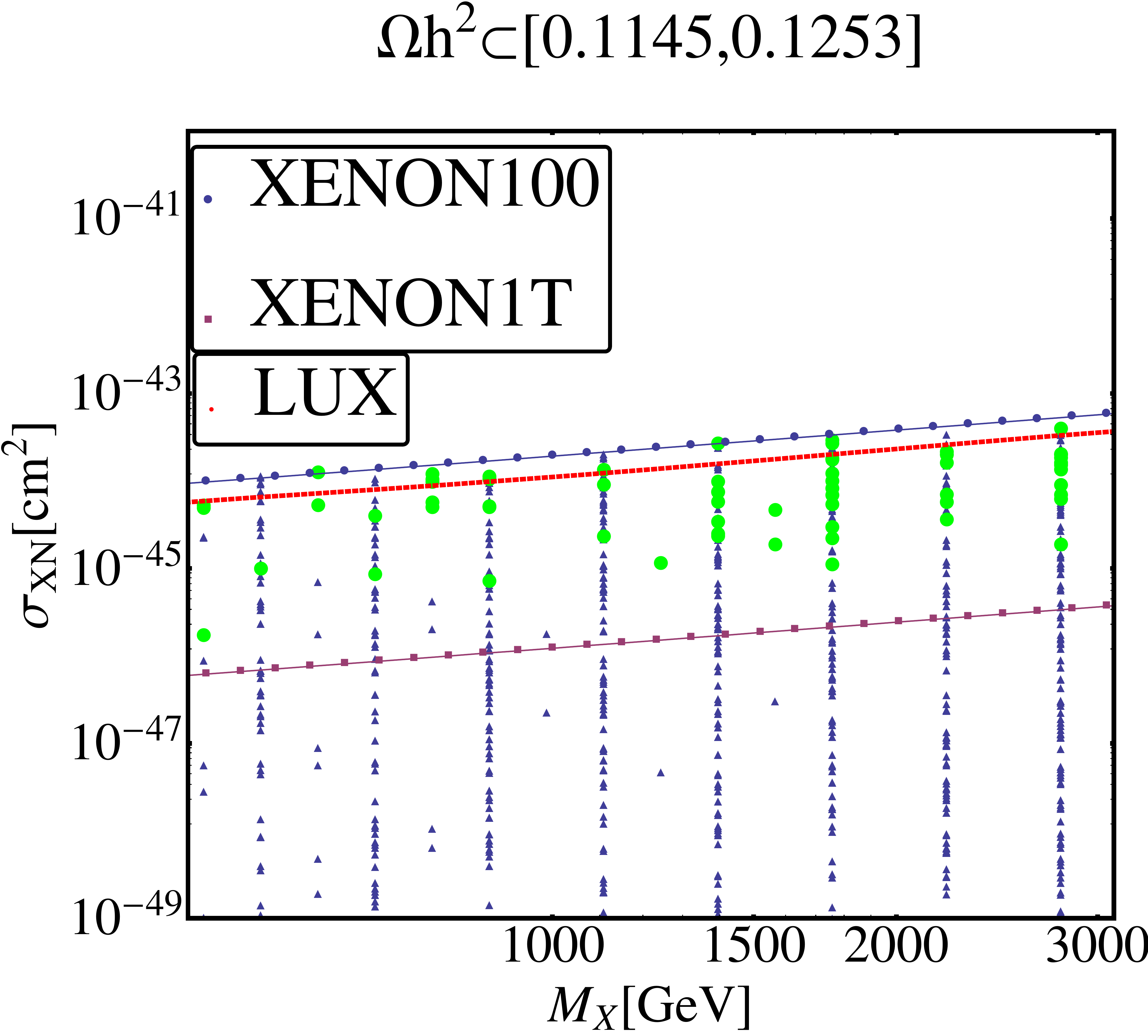}
\includegraphics[width=0.48\textwidth,height=0.462\textwidth]{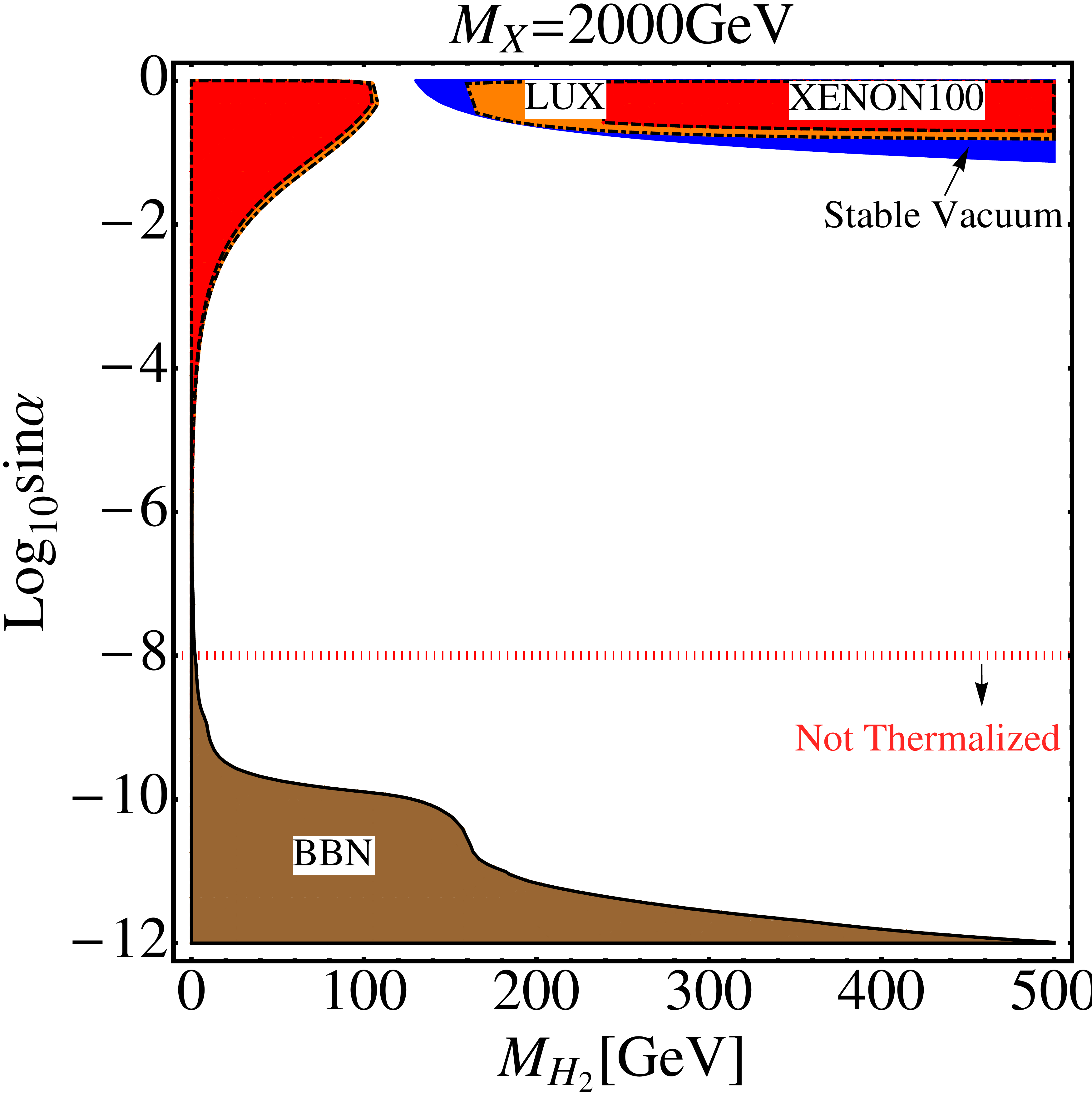}
\caption{The left panel shows the scatter plot with direct search constraints from the latest XENON100 \cite{xenon100}, LUX\cite{lux}(red line) and the future XENON1T as function as the dark matter mass and green circles give stable EW vacuum. It can be seen that all green squares can be probed by the XENON1T. The right panel shows the constraints on the mixing angle $\alpha$. The upper bound is from XENON100(red), LUX(orange) and vacuum stability(blue), and the lower bound comes from the BBN constraint on the lifetime and thermalization of $H_2$. \label{fig:xenon}}
\end{figure}

\subsection{Perturbativity and Vacuum Stability}
\begin{figure}[t]
\includegraphics[width=0.48\textwidth,height=0.462\textwidth]{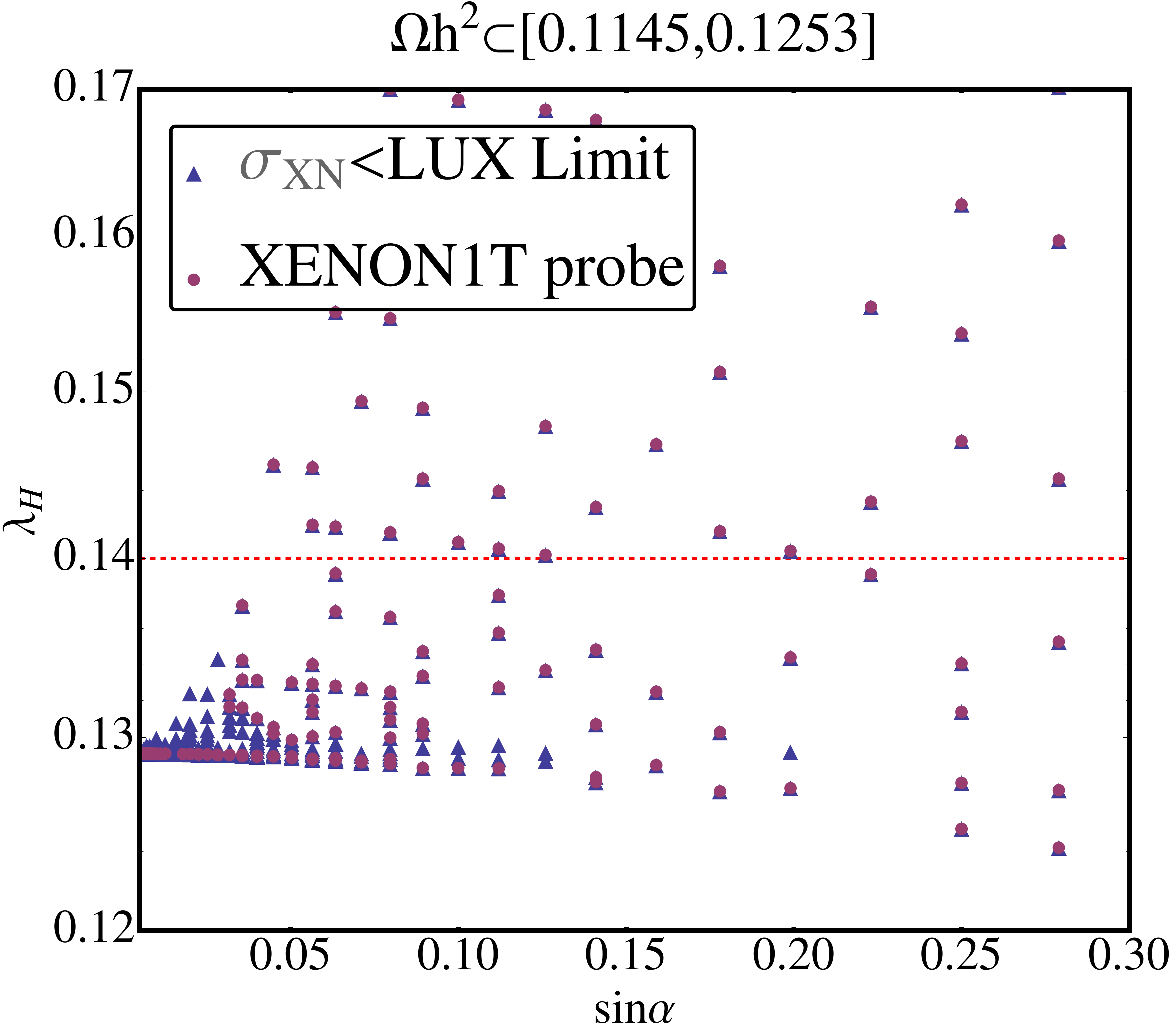}
\includegraphics[width=0.48\textwidth,height=0.462\textwidth]{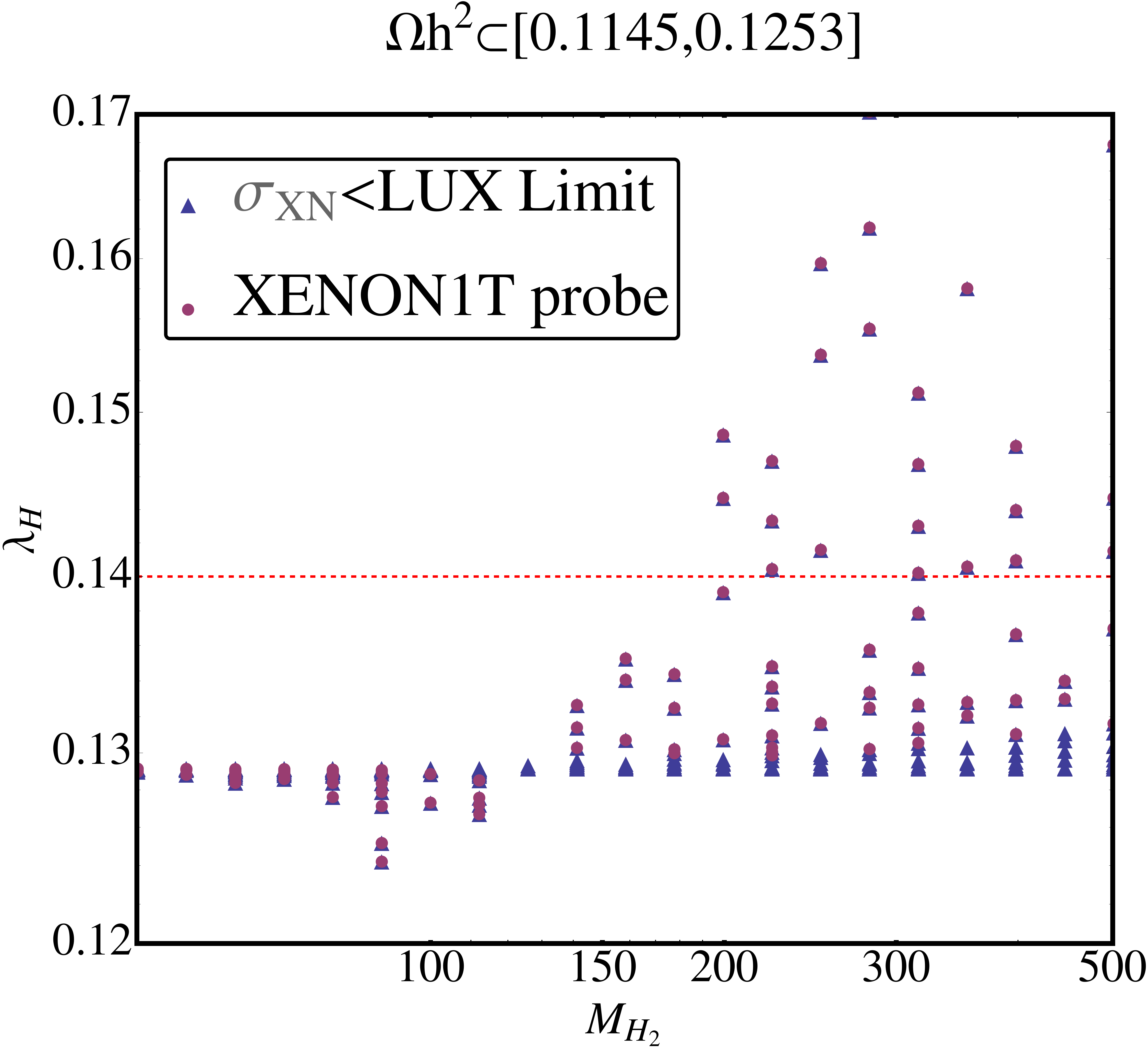}
\caption{Scatter plots show the Higgs quartic coupling $\lambda_H$ with different x-axis for $M_{X}\sim \mathcal{O}(\mathrm{TeV})$. Every point satisfies the relic density in $2\sigma$. Blue triangles are below the LUX Limit and purple circles can be probed by dark matter direct search in the near future. Regions above the horizonal red dotted line give stable EW vacuum.\label{fig:hself}}
\end{figure}

Perturbativity and vacuum stability of the model can be determined by running 
RGEs~\cite{ko_vdm} to higher energy scales: 
\begin{eqnarray*}
\frac{d\lambda_{H}}{d\ln\mu} & = & \frac{1}{16\pi^{2}}\left[24\lambda_{H}^{2}+\lambda_{H\Phi}^{2}-6y_{t}^{4}+\frac{3}{8}\left(2g_{2}^{2}+\left(g_{1}^{2}+g_{2}^{2}\right)^{2}\right)-\lambda_{H}\left(9g_{2}^{2}+3g_{1}^{2}-12y_{t}^{2}\right)\right],\\
\frac{d\lambda_{H\Phi}}{d\ln\mu} & = & \frac{1}{16\pi^{2}}\left[2\lambda_{H\Phi}\left(6\lambda_{H}+4\lambda_{\Phi}+2\lambda_{H\Phi}\right)-\lambda_{H\Phi}\left(\frac{9}{2}g_{2}^{2}+\frac{3}{2}g_{1}^{2}-6y_{t}^{2}+6g_{X}^{2}\right)\right],\\
\frac{d\lambda_{\Phi}}{d\ln\mu} & = & \frac{1}{16\pi^{2}}\left[2\left(\lambda_{H\Phi}^{2}+10\lambda_{\Phi}^{2}+3g_{X}^{4}\right)-12\lambda_{\Phi}g_{X}^{2}\right],\\
\frac{dg_{X}}{d\ln\mu} & = & \frac{1}{16\pi^{2}}\frac{1}{3}g_{X}^{3} ~.
\end{eqnarray*}
For small $\lambda_{\Phi}$ and $\lambda_{H\Phi}$, the dark sector
has negligible effects on the RG running of $\lambda_{H}$. Then similarly to the SM, 
the top quark makes a negative contribution to $\lambda_H$  from the large top Yukawa 
coupling $y_t$, and $\lambda_H$ would run to a negative value at high scale $M_\Lambda$,
leading to a metastable electroweak vacuum whose lifetime is longer than the age of our Universe. Although the precise $M_\Lambda$ depends sensitively on $y_t$ and strong coupling constant $\alpha_s$, we can use their central values and require positivity of 
$\lambda_H$ at scales larger than $10^{15}$GeV.  Then we would need 
$\lambda_H\gtrsim 0.14$ at the weak scale, and this would put a constraint on 
$M_{H_2}$ and $\sin\alpha$ from the following relation~\cite{ko_vdm}: 
\begin{equation}\label{eq:lambdaH}
 \lambda_H=\frac{M_{H_1}^2\cos^2\alpha+M_{H_2}^2\sin^2{\alpha}}{2v^2_H}\gtrsim 0.14.
\end{equation}
The allowed parameter space is shown as the blue region in the right panel of Fig.~\ref{fig:xenon} and the electroweak vacuum is metastable outside of the region. 
We also show scatter plots for $\lambda_H$ vs. $\sin\alpha$ and $\lambda_H$ vs. 
$M_{H_2}$ in Fig.~\ref{fig:hself}.   A sizable deviation from the SM value is possible 
within the current limits on $\sin\alpha$ by thermal relic density and direct search for 
$X_\mu$. 
Since the deviation can be as large as $\mathcal{O}(10\%)$ at tree level, it might be 
probed at future colliders, such as the ILC for instance.
Moreover, all points giving the stable EW vacuum can be tested at XENON1T,  
as we shall discuss in the following subsection. 

The perturbative limit is set by the input value of $g_{X}$. We find that 
$g_{X}\lesssim 1.6 (1.5) $ can give a perturbative theory up to $M_{GUT}$ (Planck scale), 
respectively.   Correspondingly, the VDM mass is  bounded from above, 
$M_{X}\simeq 7$ TeV for  $g_{X}\simeq1.5$ from Eq.~(\ref{eq:gxmx}) if nonperturbative 
effect is neglected. 

\subsection{Direct search}
The VDM $X_{\mu}$ can interact with a nucleus through the mixing of $h$ and 
$\varphi$. The cross section of $X_{\mu}$'s scattering off a nucleon is given by 
\[
\sigma\left(X_{\mu}N\rightarrow X_{\mu}N\right)=\frac{1}{16\pi}g_{X}^{4}\sin^{2}2\alpha\frac{f^{2}m_{N}^{2}}{v_{H}^{2}}\left(\frac{1}{m_{H_{2}}^{2}}-\frac{1}{m_{H_{1}}^{2}}\right)^{2}\left(\frac{M_{X}m_{N}}{M_{X}+m_{N}}\right)^{2}.
\]
Note that there is a generic cancellation between the $H_1$ and $H_2$ contributions
~\cite{ko_vdm}.
When $M_{X}\gg m_{N}$, $\frac{M_{X}m_{N}}{M_{X}+m_{N}}\simeq m_{N}$,
direct dark matter search experiments will only constrain the product
$g_{X}^{4}\sin^{2}2\alpha$, independent of $M_{X}$. In Fig.~\ref{fig:constraints},
we show the contours for 
$\sigma_{XN}\equiv\sigma\left(X_{\mu}N\rightarrow X_{\mu}N\right)$.
The solid(dashed) curve corresponds $\sigma_{XN}=10^{-44}\left(10^{-45}\right)$$\mathrm{cm}^{2}$,
region on the right-handed side gives larger $\sigma_{XN}$. Note
that for large $M_{X}$ the XENON100's bounds \cite{xenon100} are around $2\times\left(\frac{M_{X}}{1{\rm TeV}}\right)\times10^{-44}{\rm cm^{2}}$ and LUX~\cite{lux} improved the limit by a factor of 2. 

In Figs.~\ref{fig:scatter},~\ref{fig:xenon} and \ref{fig:hself}, we show the scatter  
plots for the relevant parameters which satisfy the constraints from relic density and 
LUX and can be probed by the near future XENON1T experiment. We can observe that most 
parameter space except $M_{H_2}\simeq M_{H_1}$ where the cancellation occurs or no-mixing, $\sin\alpha\simeq0$, can be covered by XENON1T. If we require the electroweak vacuum 
is stable up to high energy scale,  then all the allowed points are covered by XENON1T. 
This is explicitly shown as green squares in Fig.~\ref{fig:scatter} and green 
circles in Fig.~\ref{fig:xenon}.

\subsection{Thermalization of $X_\mu$ and $H_2$}

When  calculating  thermal relic density of VDM, we are implicitly assuming 
$X_\mu$ still has the same temperature as the thermal bath before it freezes out. 
This is justified as long as $H_2$ is in equilibrium\footnote{We require $H_2$ is in chemical equilibrium since kinetic decoupling occurs much later and give less stringent constraints.} since $X_\mu$ is thermalized by 
$X_{\mu}$-$H_2$ interaction.  In general, all the relevant processes, such as scattering one $H_2+Y\leftrightarrow H_2+Y$($Y$ is any other particle in the thermal bath), may have to 
be considered. As an illustration, in the limit of a tiny mixing angle $\alpha$, we consider 
one channel for thermalizing $H_{2}$: $H_{2}H_{2}\leftrightarrow H_{1}H_{1}$. This is 
the most efficient one for thermalization when the temperature is high. 
We then have the approximate relations: 
\[
\Gamma\simeq n_{H_{2}}\times\langle\sigma v\rangle_{H_{2}H_{2}\leftrightarrow H_{1}H_{1}},\; n_{H_{2}}\sim T^{3},\;\langle\sigma v\rangle\sim\frac{\lambda_{H\Phi}^{2}}{T^{2}}  .
\]
In the radiation dominated era, the Hubble constant is $H\sim T^{2}/M_{pl}$,
so the condition for equilibrium gives 
\[
\Gamma\gtrsim H\Rightarrow \lambda_{H\Phi}^{2}\gtrsim \frac{T}{M_{pl}}.
\]
For $T\simeq 1$TeV we have $|\lambda_{H\Phi}|\gtrsim 10^{-8}$, which in turn leads 
to $\sin\alpha\gtrsim10^{-8}$.  in the right panel of Fig.~\ref{fig:xenon}, we show 
the region under the red dotted line where the dark sector is not thermalized.   
This constraint is much more stringent than BBN constraint except in the very 
low $M_{H_2}$ region. 
The main purpose of the above discussion is  to show that even for very tiny 
$\lambda_{H\Phi}$, $H_{2}$ can be still in thermal equilibrium when $X_{\mu}$ starts 
to freeze out.

\section{Indirect Signatures from VDM pair annihilation}\label{sec:anniDM}
\begin{figure}[t]
\includegraphics[scale=0.5]{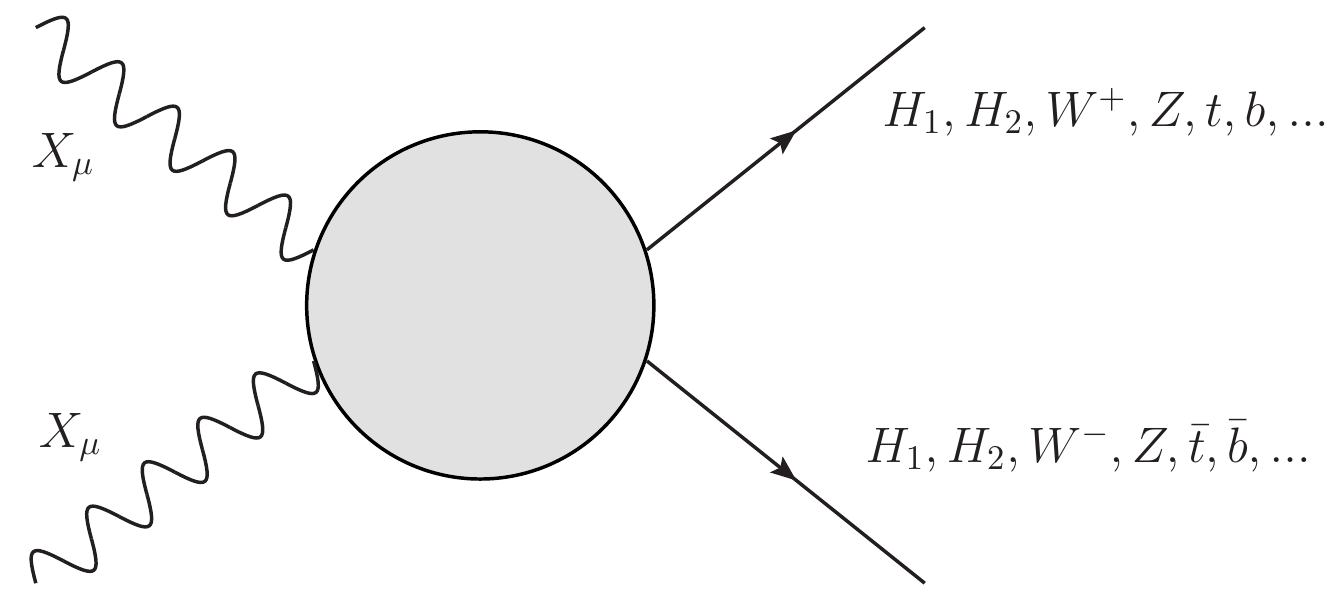}
\caption{annihilation process.\label{fig:anni}}
\end{figure}
Most phenomenology of the VDM with mass $\mathcal{O}(100\mathrm{GeV})$ has been 
studied in Ref.~\cite{ko_vdm}, except for its indirect signatures. 
In this section, we focus on indirect signatures from the pair annihilation of VDM 
(as shown in Fig.~\ref{fig:anni}) which are described by the renormalizable model 
Lagrangian,  Eq.~(2.1). Depending on the parameters, the dominant annihilation 
channels can be different, resulting in different spectra for cosmic rays. 

Since it is impractical to show all the cases, here we only discuss 4 different 
cases, as tabulated in Table~\ref{tab:benchmarks}.   
For a TeV VDM ($X_\mu$),  the reaction $X_\mu X_\mu \rightarrow H_2H_2$ is the 
dominant annihilation channel.   Therefore the spectrum shape will be truncated at 
$M_X$,  and fully determined by decay modes of $H_2$. The lighter $X_\mu$ cases, C 
and D, are chosen just for completeness and comparison with the cases A and B. 
All these cases are still allowed by current experimental constraints considered in the 
previous section, giving the correct thermal relic density of the VDM, although the 
dominant annihilation channels for indirect detection for each case could be quite 
different from each other.   Note that the case B would not give an absolute stable vacuum 
but a metastable vacuum,  and here we choose this low $M_{H_2}$ case just for comparison.

The production rate for cosmic rays from DM pair annihilation is given by 
\cite{Bertone:2010zza}
\begin{equation}
Q\left(E,\vec{r}\right)=\frac{1}{2}\left(\frac{\rho\left(\vec{r}\right)}{M_{\mathrm{DM}}}\right)^2 \sum_i \langle \sigma v\rangle_i\frac{dN_i}{dE}.\label{eq:source}
\end{equation}
$\frac{dN_i}{dE}$ is the energy spectrum function from a specific annihilation channel 
$i$, and $M_{\mathrm{DM}}=M_{X}$ in our discussion.  The function $\rho\left(r\right)$ 
is the density profile of dark matter. We shall use the Navarro-Frenk-White (NFW) density 
profile~\cite{nfw},
\[
\rho\left(\vec{r}\right)=\rho_{\odot}\left[\frac{r_{\odot}}{r}\right]\left[\frac{1+\left(r_{\odot}/r_{c}\right)}{1+\left(r/r_{c}\right)}\right]^{2}.
\]
Here we use the default values in \texttt{micrOMEGAs-3.1}: 
$\rho_{\odot}\simeq0.3\mathrm{GeV/cm^{3}},\; 
r_{\odot}\simeq8.5\mathrm{kpc}$  and $r_{c}\simeq20\mathrm{kpc}$ ~\cite{micromegas}. 
After production, charged particles propagate  through the Galaxy and may lose part of their 
energy before reaching the solar system. Then the number density  
$\psi\left(E,r_{\odot}\right)$ can be expressed as~\cite{micromegas24}
\[
\psi\left(E,\vec{r}_{\odot}\right)=\int_{E}^{M_{X}}dE'\int d^{3}\vec{r}\; G\left(\vec{r}_{\odot},E;\vec{r},E'\right)Q\left(E',\vec{r}\right),
\]
where $G\left(\vec{r}_{\odot},E;\vec{r},E'\right)$ is the Green's function, paremetrizing the 
effect during the propagation, such as diffusion and energy loss. 
Finally the flux is given by 
\[
\Phi=\frac{v\left(E\right)}{4\pi}\psi.
\]

For $\gamma$-ray and neutrinos, after production they travel almost freely, so the fluxes 
are only dependent on angle region of observation and the integral of squared $\rho$ over the line of sight
\[
\Phi_{\gamma / \nu} \propto 2\pi \int \sin\theta d\theta \int_0^{\infty}dr \rho ^2 (r'),
\]
where $r'=\sqrt{r^2+r^2_\odot -2 r r_\odot \cos \theta}$ and $\theta$ is the angle between 
the line of sight and the center of Milky Way. We shall use $\theta=\pi/6$  
(integrating the region with $\delta \theta = \pi/60$) as an example and neglect the 
$\gamma$-ray  induced by inverse Compton scattering and synchrotron radiation from 
the primary $e^+$  and $\bar{p}$ for simplicity. This is justified as long as we concentrate on the high energy part of the spectrum from $10^{-2}M_X$ to $M_X$. Relative sizes of various contributions are illustrated in~\cite{Ibarra:2009dr}. To calculate the cosmic-ray spectra 
from VDM pair annihilation (i.e. positrons, antiprotons, gammas and neutrinos), we have used \texttt{micrOMEGAs-3.1}~\cite{micromegas} which used \texttt{Pythia}~\cite{pythia} inside.

\begin{table}[tb]
\begin{tabular}{|c|c|c|c|c|c|c|}
\hline
 \parbox[c]{2.5cm}{{--}} & $M_X$[GeV] & $M_{H_2}$[GeV] & \parbox[c]{1.2cm}{$g_X$} & \parbox[c]{1.2cm}{$\sin{\alpha}$} & $ \langle \sigma v \rangle$[$10^{-26}\mathrm{cm}^3$/s] & $\sigma _{XN}[10^{-45}\mathrm{cm}^2]$ \\
\hline
case A & 1100 & 280& 0.56 & 0.08 & 2.25 & 0.77 \\
\hline
case B & 1100 & 90 & 0.56 & 0.05 & 2.36 & 0.40 \\
\hline
case C & 400 & 500 & 0.34 & 0.30 & 2.32 & 2.68 \\
\hline
case D & 400 & 250 & 0.37 & 0.14 & 2.35 & 0.46 \\
\hline
\end{tabular}
\caption{Four cases for illustrating indirect signatures, all are still allowed by current experimental constraints.\label{tab:benchmarks}}
\end{table}

\begin{figure}
\includegraphics[width=0.49\textwidth,height=0.33\textwidth]{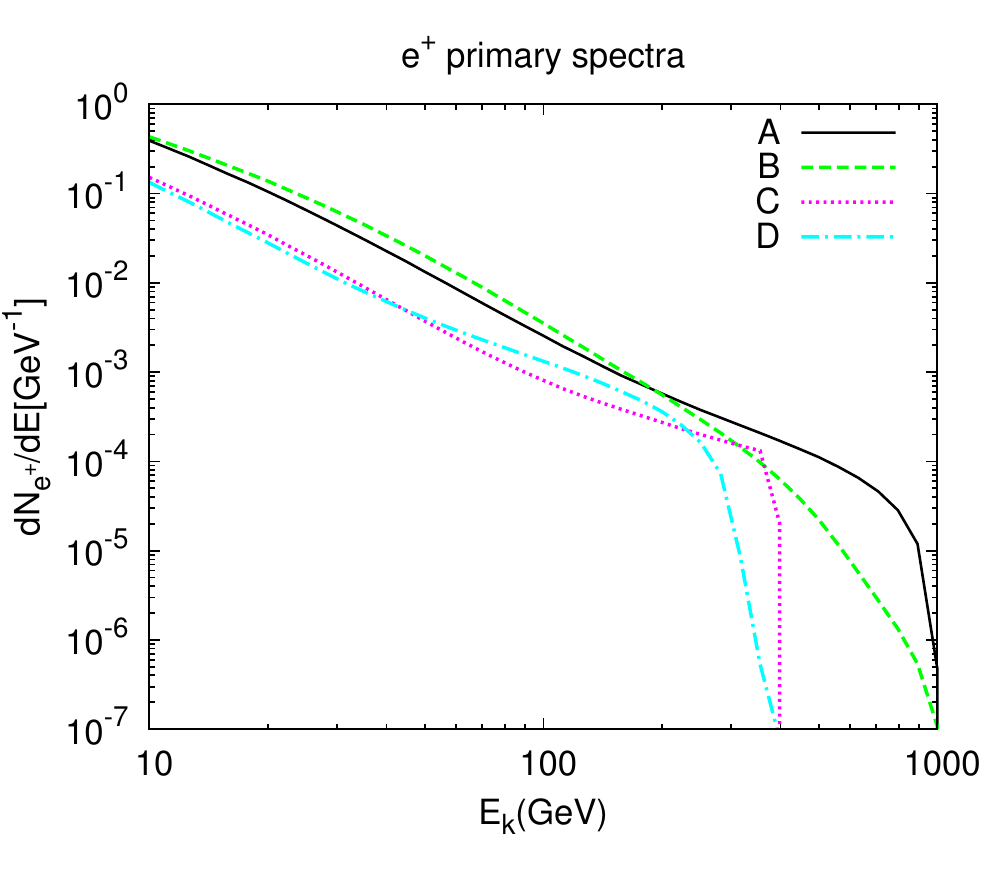}
\includegraphics[width=0.49\textwidth,height=0.33\textwidth]{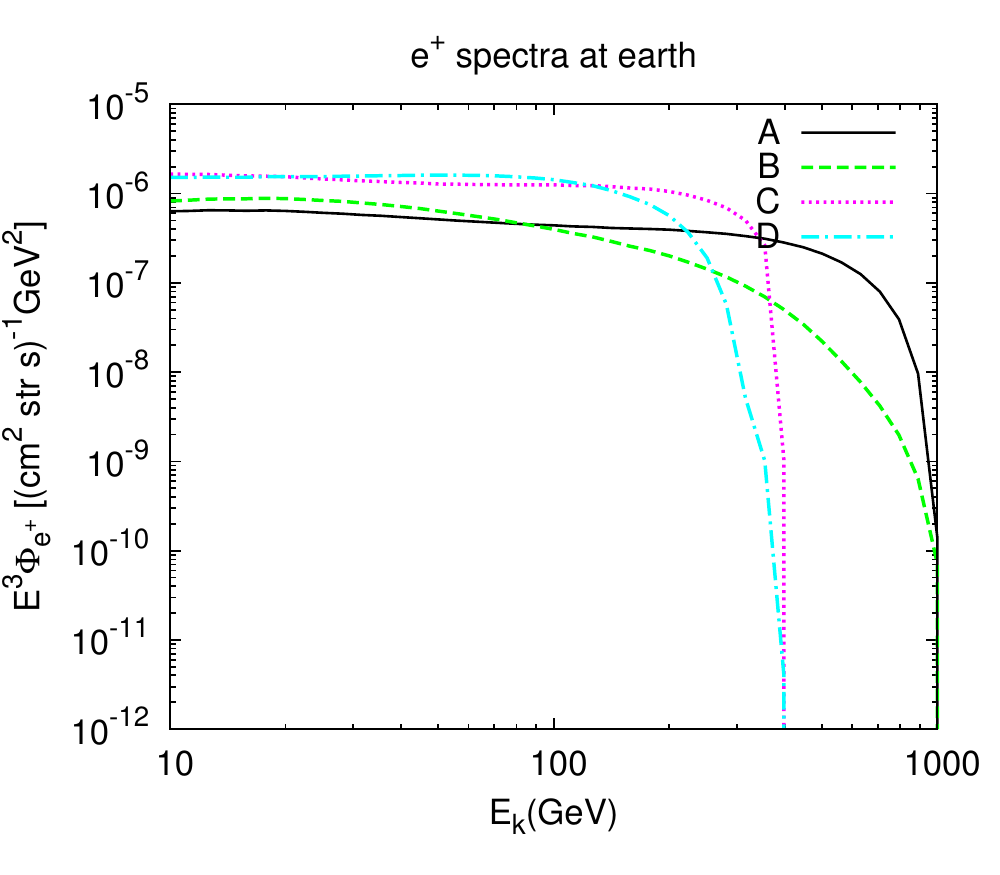}
\includegraphics[width=0.49\textwidth,height=0.33\textwidth]{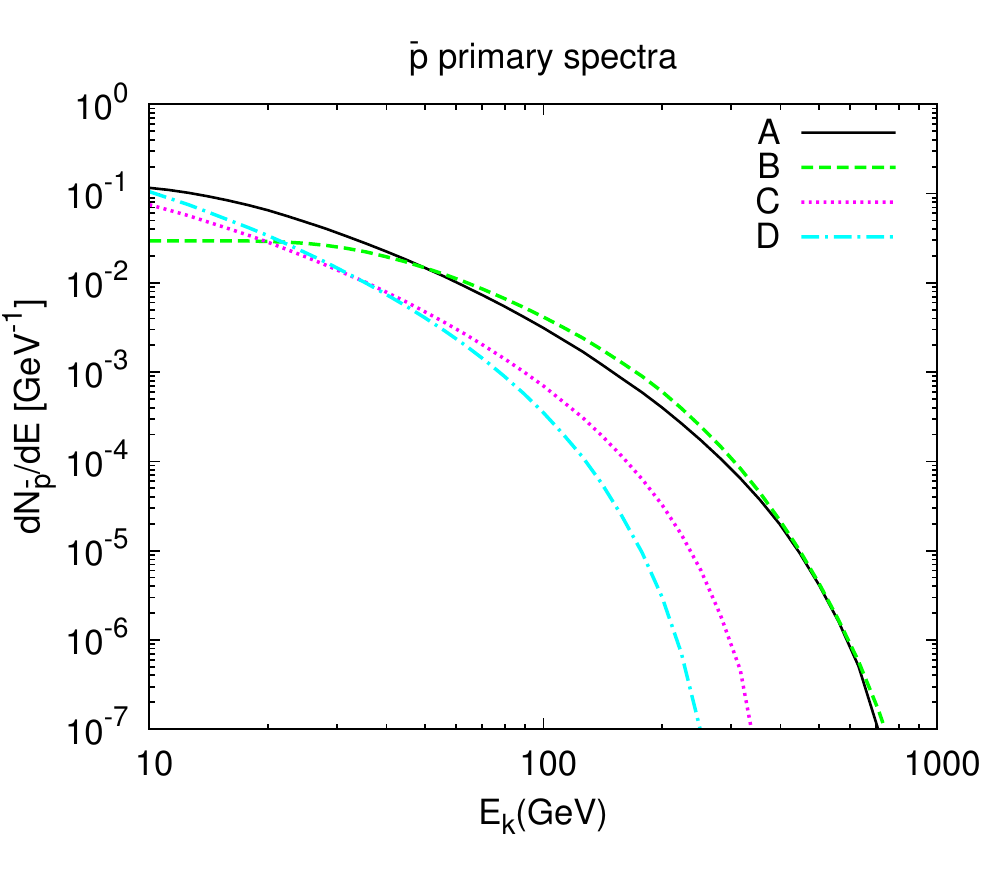}
\includegraphics[width=0.49\textwidth,height=0.33\textwidth]{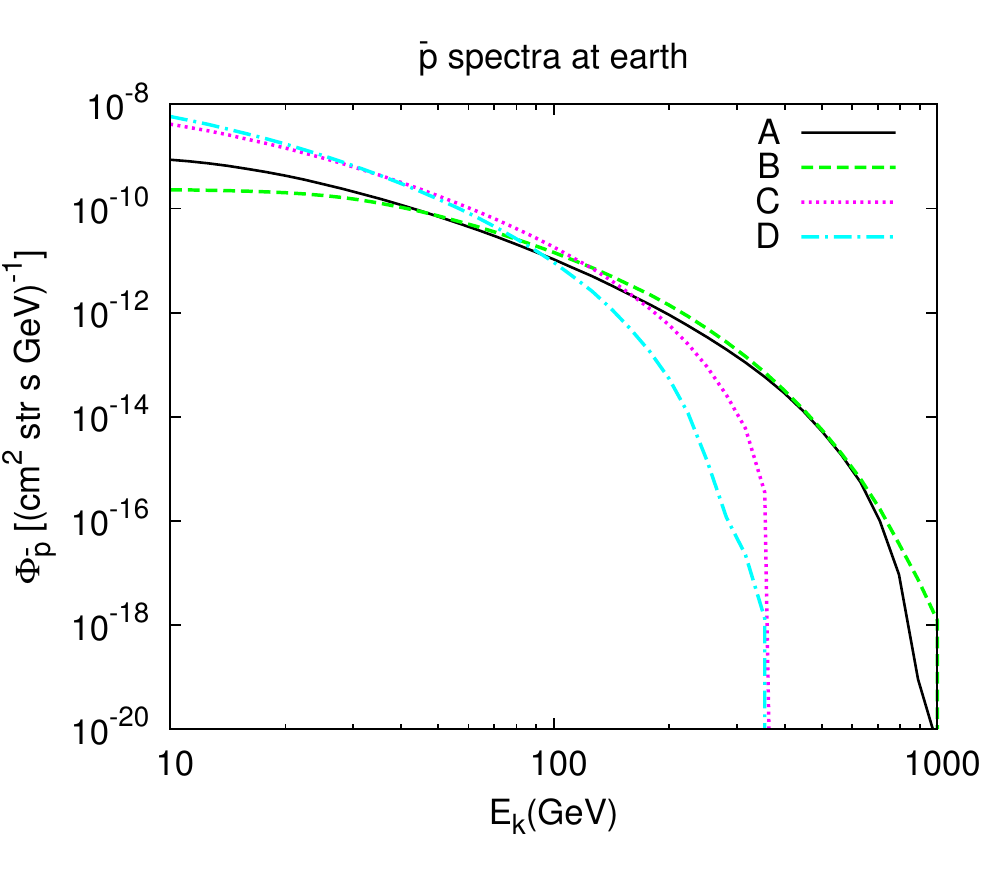}
\includegraphics[width=0.49\textwidth,height=0.33\textwidth]{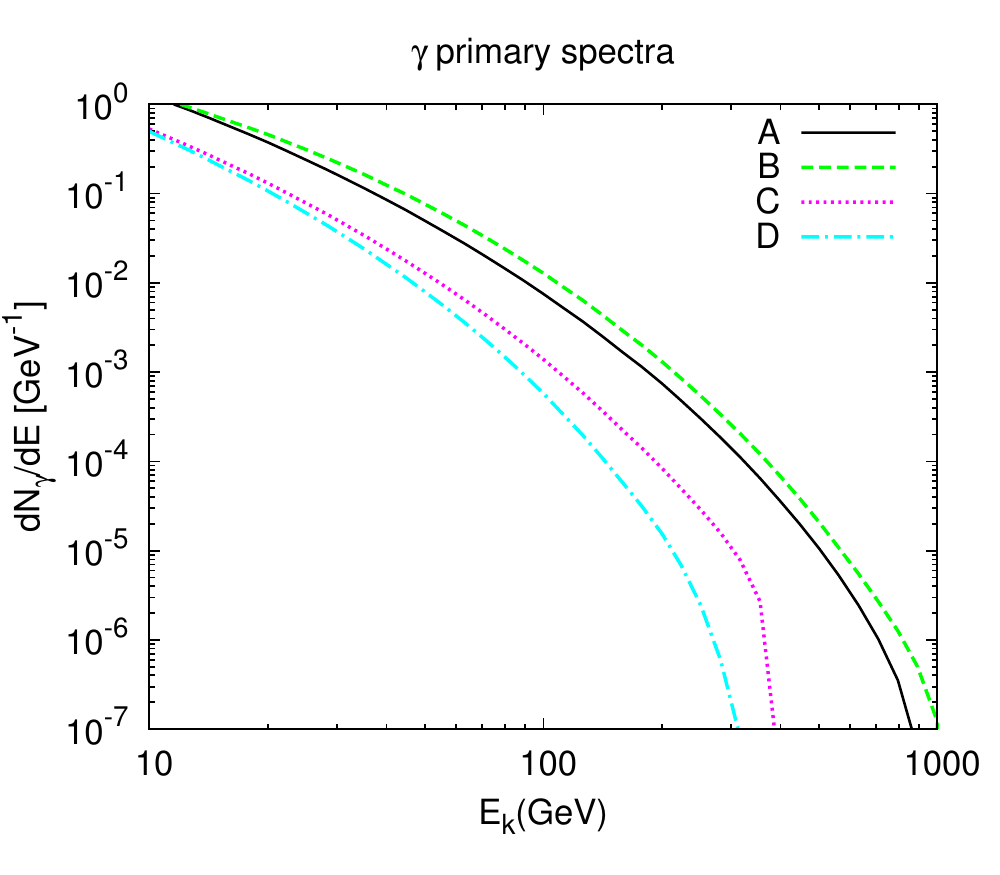}
\includegraphics[width=0.49\textwidth,height=0.33\textwidth]{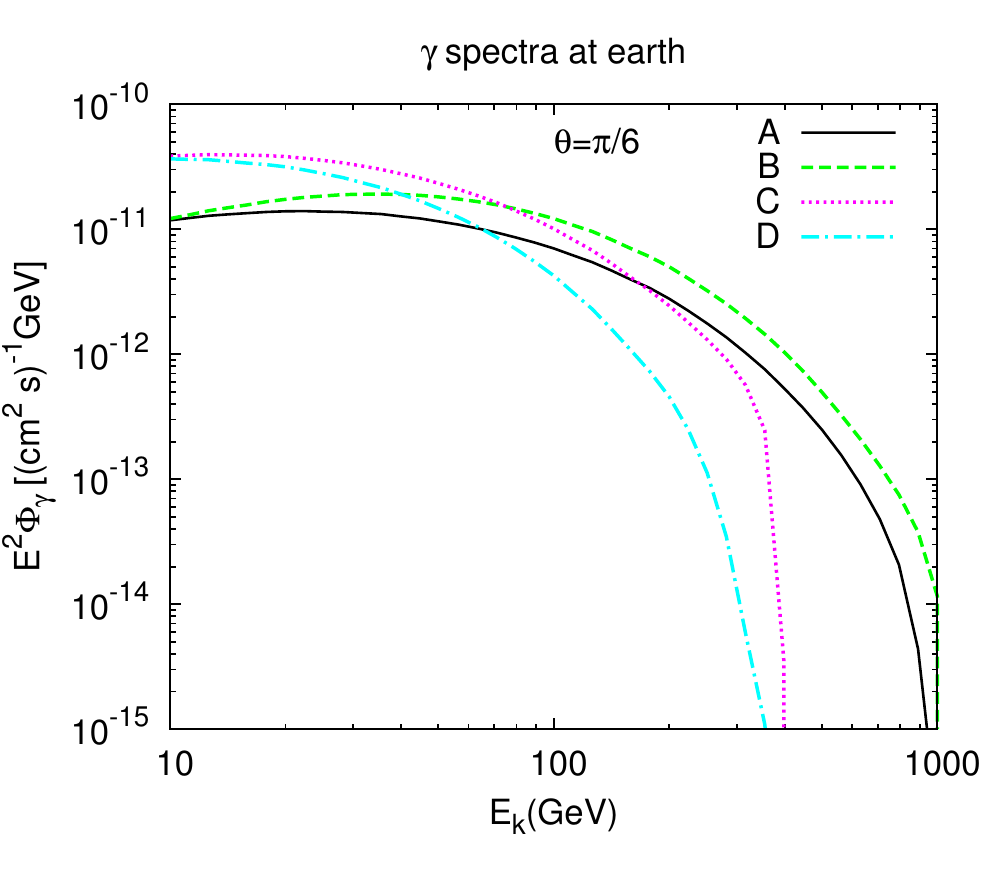}
\includegraphics[width=0.49\textwidth,height=0.33\textwidth]{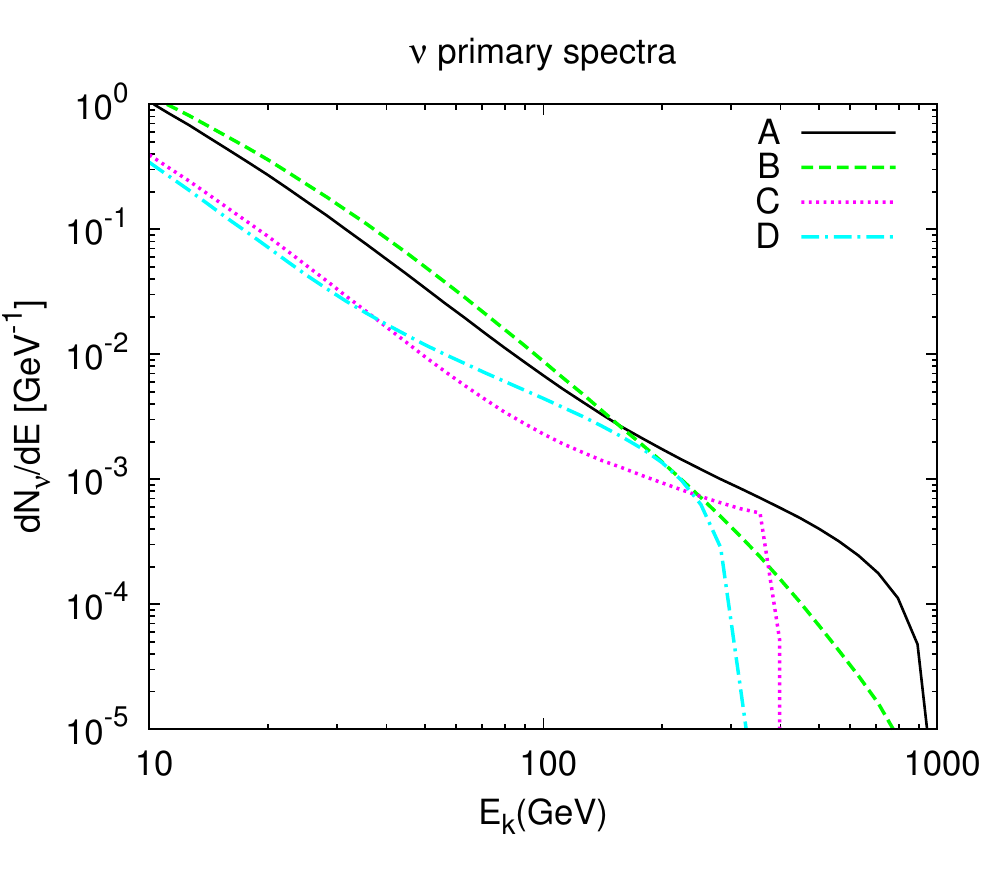}
\includegraphics[width=0.49\textwidth,height=0.33\textwidth]{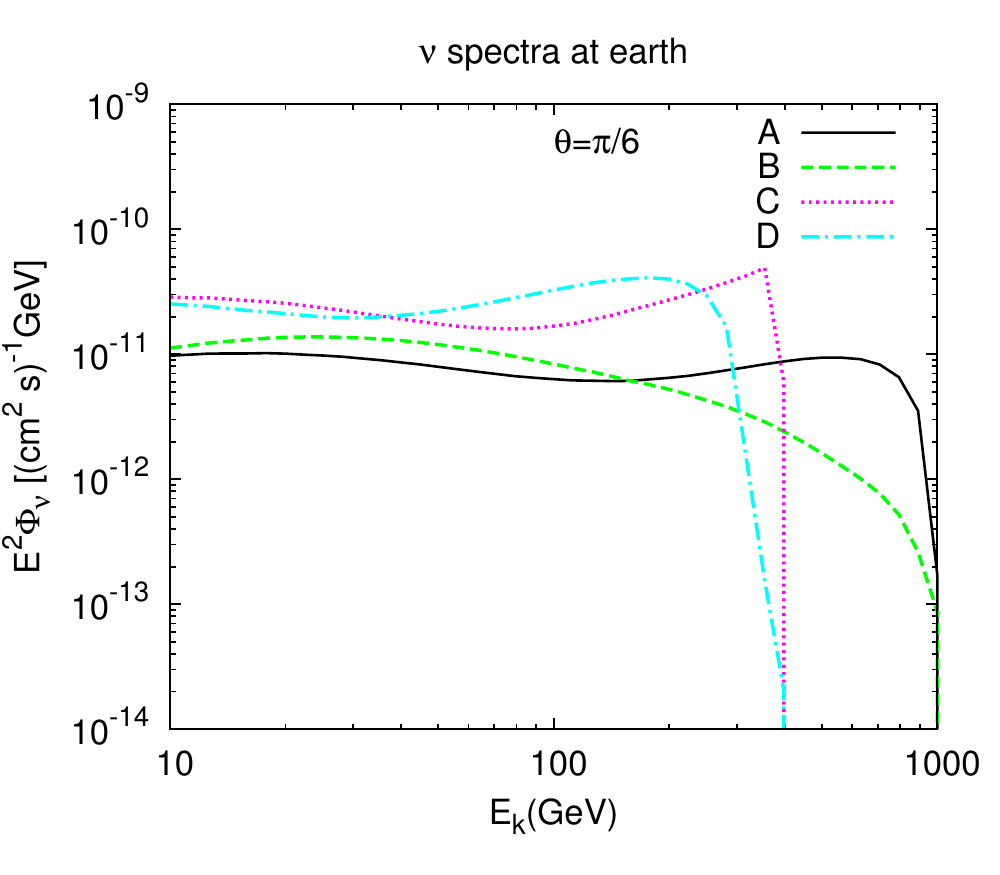}
\caption{Spectra of $e^+$, $\bar{p}$, $\gamma$ and $\nu$ from vector dark matter annihilation only. The left panels show the primary spectra while the right ones show the spectra at earth after propagation(the units are chosen with the usual convention).\label{fig:annispectrum} }
\end{figure}

Generally, the VDM mass determines the energy cut-off of the primary cosmic ray spectra. 
Since $H_2$ couples to the SM particles with the same pattern as that of $H_1$, its mass 
determines the branching ratios completely. These different decay final products, together 
with relevant importance of annihilation channels, can lead different spectra for cosmic 
rays, as shown in Fig.~\ref{fig:annispectrum}, although all of them have similar size of  
$\langle \sigma v \rangle$.

For instance in $e^+$ spectrum, in the case C, the dominant annihilation channel is 
$X + X \rightarrow W^+ + W^-$, while $X + X \rightarrow H_2 + H_2$ 
is the dominant one for the case D. About $1/3$ of $W$ decay directly to charged leptons and 
the rest decay hadronically, so that in sum  the multiplicity for charged particles
is about $20$ in a single W decay~\cite{pdg}.  While a $250$GeV $H_2$ mostly decays to $ZZ$ and 
$W^+W^-$ whose decay products are then boosted differently. This is the main reason for the different spectra in the case C and D. Similar mechanisms apply to other spectra for $\gamma$, 
$\bar{p}$ and $\nu$'s. For the overall differences between the case A/B and the case C/D, 
the fluxes at earth have the opposite behavior in some energy ranges, although the primary 
spectra $dN/dE$ in case A and B are larger than those in case C and D. 
This is mainly due to the factor $\left(\rho/M_{\mathrm{DM}}\right)^2 $ in the source 
function $Q$, Eq.~(\ref{eq:source}), and lighter dark matter tends to have larger flux 
$\Phi$ in the kinematically allowed energy range. 

Note that the spectra we discussed so far are only the signatures from dark matter 
pair annihilation. 
In reality astrophysical observations of cosmic rays and $\gamma$-ray are the sum 
of a much larger backgrounds and the above signals.
For instance, the positron ($e^+$) flux  with $10\textrm{GeV}<E_k<300\textrm{GeV}$ 
from VDM annihilation in Fig.~\ref{fig:annispectrum} has $E^3\Phi_{e^+}$ around 
$10^{-6}(\textrm{cm}^{2}\textrm{str s})^{-1}\textrm{GeV}^{2} $, while the background $E^3\Phi_{e^+}$ is about $10^{-3}(\textrm{cm}^{2}\textrm{str s})^{-1}\textrm{GeV}^{2}$ 
which can be inferred from the data in Fig.~\ref{fig:cosmicray} of the next section. 
Therefore,  for canonical values of thermal $\langle \sigma v\rangle$, 
the differences among those cases can hardly be distinguished 
unless there are some mechanisms for boosting the spectrum, such as Sommerfeld 
enhancement \cite{Hisano:2006nn,ArkaniHamed:2008qn} or Breit-Wigner resonance 
\cite{Hisano:2004ds,Ibe:2008ye,Guo:2009aj} enhancement for explaining the recent 
observed positron excesses in~\cite{pamela,fermilat,ams02}. However, stringent constraints from CMB have been put on such mechanisms for annihilating dark matter~\cite{Padmanabhan:2005es, Galli:2009zc, Slatyer:2009yq, Zavala:2009mi, Madhavacheril:2013cna} \footnote{Although possible exceptions exist~\cite{Hannestad:2010zt, Vincent:2010kv, Finkbeiner:2010sm, Cirelli:2010nh}, such as multi-component dark matter, halo substructure.}. 
Therefore we conclude that the VDM pair annihilations from the renormalizable Lagrangian 
(2.1) has difficulties to explain the positron excess observed by PAMELA and AMS02. 
And we shall focus on the decaying dark matter scenario in the next section.    

\section{Indirect signatures from VDM ($X_\mu$) decay}
\label{sec:higher}
\subsection{Effective Operators for decaying VDM ($X_\mu$) }
In the renormalizable theory described by the Lagrangian (2.1), 
the dark matter $X_{\mu}$ can not decay because of the $Z_2$ symmetry we assumed.
This is not true any more if higher dimensional nonrenormalizable operators are taken 
into  account.   Generally, higher dimensional operators are suppressed by the power of 
some new physics scale $\Lambda$, above which the nonrenormalizable Lagrangian 
begins to violate unitarity and fails to describe physical phenomena correctly. 
In the absence of a complete theory above $\Lambda$, we may write down all  
the operators which are invariant under the gauge group $G_{SM}\times U(1)_{X}$.
To which orders we shall truncate is dependent on the observable we  are considering. 
In the following, we only list higher dimensional operators up to dim-6, especially 
we focus on those involving both fields from the dark sector and from the SM sector. 
Such higher dimensional operators for the SM sector upto dim-6 can be found in Ref.s~
\cite{Buchmuller:1985jz,Grzadkowski:2010es}. 

Since there are two fields $\Phi$ and $X^{\mu}$ in the dark sector, 
gauge invariant operators in the dark sector would be made of the following operators: 
\[
\Phi^{\dagger}\Phi,\;\Phi^{\dagger}i\overleftrightarrow{D}_{\mu}\Phi,\; X^{\mu\nu},
\;\tilde{X}^{\mu\nu}.
\]
where $\Phi^{\dagger}\overleftrightarrow{D}_{\mu}\Phi=
\Phi^{\dagger}D_{\mu}\Phi-\left(D_{\mu}\Phi\right)^{\dagger}\Phi$.
The independent effective operators of dim-6 in this sector  are
\[
\left(\Phi^{\dagger}\Phi\right)^{3},\;\left(\Phi^{\dagger}\Phi\right)\square\left(\Phi^{\dagger}\Phi 
\right),\;\left(\Phi^{\dagger}D^{\mu}\Phi\right)^{\dagger}\left(\Phi^{\dagger}D^{\mu}\Phi\right),\Phi^{\dagger}\Phi X_{\mu\nu}X^{\mu\nu},\;\Phi^{\dagger}\Phi\tilde{X}_{\mu\nu}X^{\mu\nu}.
\]
The operator $\left(D_{\mu}\Phi\right)^{\dagger}\left(D_{\nu}\Phi\right)X^{\mu\nu}$
is redundant, as it can be shown by partial integration and using equations
of motion. 

Gauge invariant operators in SM sector are products of the following: 
\[
H^{\dagger}H,\; H^{\dagger}i\overleftrightarrow{D}_{\mu}H,\; B^{\mu\nu},\;\tilde{B}^{\mu\nu},\;\bar{L_{i}}R_{j}H,\;\bar{f}_{i}\gamma^{\mu}f_{j},\;\left(\bar{L}_{i}\sigma^{\mu\nu}R_{j}\right)H,\; H^{\dagger}\tau^{I}HW_{\mu\nu}^{I},\; H^{\dagger}\tau^{I}H\tilde{W}_{\mu\nu}^{I},
\]
where $L$ and $R$ stand for left-handed and right-handed fermion
fields, respectively.  Note that there is only one dimension-five operator within the 
SM sector, namely the Weinberg operator for Majorana neutrino masses.  

Dimension-6 operators that involve both SM and dark sector fields
are
\begin{eqnarray*}
 &  & \left(\Phi^{\dagger}\Phi\right)^{2}H^{\dagger}H,\;\Phi^{\dagger}\Phi\left(H^{\dagger}H\right)^{2},\Phi^{\dagger}\Phi\square H^{\dagger}H,\;\left(\Phi^{\dagger}i\overleftrightarrow{D}_{\mu}\Phi\right)\left(H^{\dagger}i\overleftrightarrow{D}_{\mu}H\right),\\
 &  & \Phi^{\dagger}\Phi\left(\bar{L_{i}}R_{j}H+h.c\right),\;\left(\Phi^{\dagger}i\overleftrightarrow{D}_{\mu}\Phi\right)\left(\bar{L}_{i}\gamma^{\mu}L_{j}+\bar{R}_{i}\gamma^{\mu}R_{j}\right),\;\left(\bar{L}_{i}\sigma_{\mu\nu}R_{j}\right)HX^{\mu\nu}+h.c,\\
 &  & \Phi^{\dagger}\Phi B_{\mu\nu}X^{\mu\nu},\;\Phi^{\dagger}\Phi\tilde{B}_{\mu\nu}X^{\mu\nu},H^{\dagger}HB_{\mu\nu}X^{\mu\nu},\; H^{\dagger}H\tilde{B}_{\mu\nu}X^{\mu\nu},\; H^{\dagger}HX_{\mu\nu}X^{\mu\nu},\; H^{\dagger}H\tilde{X}_{\mu\nu}X^{\mu\nu},\\
 &  & H^{\dagger}\tau^{I}HW_{\mu\nu}^{I}X^{\mu\nu},\; H^{\dagger}\tau^{I}H\tilde{W}_{\mu\nu}^{I}X^{\mu\nu}.
\end{eqnarray*}
The above operators make the whole independent set of operators with both the SM
fields and the dark sector fields.  Others can be reduced to linear combinations of these 
operators by using equations of motion.

After the spontaneous gauge symmetry breaking of $G_{SM}\times U(1)_{X}$, some of the
above effective operators can lead to dark matter $X_{\mu}$ decay.
Let us consider the following operators for the VDM decays into two SM particles 
in the final states: 
\begin{eqnarray*}
1. & \left(\Phi^{\dagger}i\overleftrightarrow{D}_{\mu}\Phi\right)\left(H^{\dagger}i\overleftrightarrow{D}^{\mu}H\right) & \Rightarrow X^{\mu}\rightarrow\varphi/h+\gamma/Z,\\
2. & \left(\Phi^{\dagger}i\overleftrightarrow{D}_{\mu}\Phi\right)\left(\bar{f}\gamma^{\mu}f\right),\;\bar{L}\sigma_{\mu\nu}RHX^{\mu\nu}+h.c & \Rightarrow X^{\mu}\rightarrow\bar{f}+f,\\
3. & \Phi^{\dagger}\Phi B_{\mu\nu}X^{\mu\nu},\;\Phi^{\dagger}\Phi\tilde{B}_{\mu\nu}X^{\mu\nu},\left(\Phi\rightarrow H\right) & \Rightarrow X^{\mu}\rightarrow \varphi/h+\gamma/Z,\\
4. & H^{\dagger}\tau^{I}HW_{\mu\nu}^{I}X^{\mu\nu},\; H^{\dagger}\tau^{I}H\tilde{W}_{\mu\nu}^{I}X^{\mu\nu} & \Rightarrow X^{\mu}\rightarrow \varphi/h+\gamma/Z,
\end{eqnarray*}
There are also some interesting three-body decay channels, such as
\[
\Phi^{\dagger}\Phi B_{\mu\nu}X^{\mu\nu}\Rightarrow X^{\mu}\rightarrow \varphi+\varphi+\gamma/Z.
\]
Generally, three-body decays from these operators are suppressed more compared 
with two-body decay because of the smaller phase space available. 
Therefore we will mainly discuss the two-body decay in the following.

\subsection{A simple UV completion}
It should be pointed out that not all of the above operators need to be investigated 
simultaneously for the purpose of the positron excess observed by PAMELA and AMS02. 
The choice is highly dependent on the exact theory beyond energy scale $\Lambda$ 
and low energy observables  we are interested in. 
For instance, Refs.~\cite{Arina:2009uq,Gustafsson:2013gca} investigated $\gamma$-ray 
in a similar framework.   Here as a concrete illustration for fermionic final states, let us 
consider the following operator 
\[
\left(\Phi^{\dagger}i\overleftrightarrow{D}_{\mu}\Phi\right)\left(\bar{f}\gamma^{\mu}f\right),
\]
which can induce a decay 
\[
X^{\mu}\rightarrow f\;\bar{f}.
\]
This operator can be induced from the following interactions when
both $\Phi$ and $f$ are charged under a new extra $U(1)'$ symmetry with $A_\mu^{'}$ 
gauge field,
\[
\mathcal{L}=\left(D'_{\mu}\Phi\right)^{\dagger}D'^{\mu}\Phi+\bar{f}i\gamma^{\mu}D'_{\mu}f-\frac{1}{4}F'^{\mu\nu}F_{\mu\nu}'+\left(D'_{\mu}\phi\right)^{\dagger}D'^{\mu}\phi-V\left(\phi^{\dagger}\phi\right),
\]
 where the covariant derivatives are
\begin{eqnarray*}
D'_{\mu}\Phi & = & \left(\partial_{\mu}+ig_{X}Q_{X}X_{\mu}+ig'Q'_{\Phi}A'_{\mu}\right)\Phi,\\
D'_{\mu}\phi & = & \left(\partial_{\mu}+ig'Q'_{\phi}A'_{\mu}\right)\phi,\\
D'_{\mu}f & = & \left(D_{\mu}^{{\rm SM}}+ig'Q'_{f}A'_{\mu}\right)f.
\end{eqnarray*}
A new scalar $\phi$ has been introduced in order to break $U(1)'$ spontaneously and 
make $A'_{\mu}$ massive. If only leptons have  $U(1)'$ charges among the SM particles, 
then the massive VDM $X^{\mu}$ would decay to a lepton pairs only,
\[
X^{\mu}\rightarrow l^{+}l^{-},\; l=e,\;\mu,\;\tau .
\]
In such a case, $U(1)'$ charge can be identified as lepton number
~\footnote{We ignore the anomaly cancellation issue in this paper.},
and $\phi$ could also couple to right-handed neutrino and give the
Majorana mass term after $U(1)'$ breaking, acting as the source of
type-I seesaw mechanism. If only $e^{\pm}$ and $\nu_{R}$ are $U(1)'$-charged,
then $X_{\mu}$ only decays to $e^{+}e^{-}$, see refs.\cite{Baek:2008nz,Bi:2009uj} for similar models. For simplicity, we shall assume $100\%$ of $X_{\mu}$ decay to a single channel for indirect
signatures.

In order to  explain the positron excess correctly, the lifetime of dark matter should be 
around $\tau_{DM}\sim10^{26}\mathrm{s}$, which determines the scale $\Lambda$: 
\[
\Gamma\sim\frac{g_{\Lambda}^{4}M^{5}}{\Lambda^{4}},\;\tau=\frac{\hbar}{\Gamma}\sim10^{26}\mathrm{s}\Rightarrow\Gamma\sim6\times10^{-51}\mathrm{GeV}.
\]
For $M=1$TeV, we have 
\[
\Lambda\sim g_{\Lambda}\left(\frac{M^{5}\tau}{\hbar}\right)^{\frac{1}{4}}=g_{\Lambda}\left(\frac{10^{15}\mathrm{GeV^{5}}\times10^{26}\mathrm{s}}{6.583\times10^{-25}\mathrm{GeV\; s}}\right)^{\frac{1}{4}}\sim2g_{\Lambda}\times10^{16}\mathrm{GeV,}
\]
If $g_{\Lambda}\sim0.1$ then $\Lambda\sim2\times10^{15}$ GeV. 
In   the framework of the above $U(1)'$ model, we have the following identifications:  
$\Lambda\rightarrow M_{A'}$,
$g_{\Lambda}\rightarrow g'$ and $M^{5}\rightarrow M_{X}^{3}v_{\Phi}^{2}$.

Note that those new nonrenormalizable interactions would not affect the VDM annihilation 
in Sec.~III or other results derived from the renormalizable part of the VDM Lagrangian, 
because the new particles are simply too heavy $\sim 10^{16}$GeV.  
Only if there were term like $\lambda_{\phi H} \phi^\dagger\phi H^\dagger H$ 
in the potential $V$, the running of $\lambda_H$ above scale $\Lambda $ will get additional 
contribution from $\phi$ and vacuum stability condition \ref{eq:lambdaH} will be modified.
This would be highly dependent on the size of $\lambda_{\phi H}$ and beyond our discussion in this paper  (see Ref.~\cite{Tang:2013bz} for a brief review). In this paper we simply assume
$\lambda_{\phi H}$ term is negligibly small. 

\subsection{Decaying VDM ($X_\mu$) and positron excesses}\label{sec:excess}
Discussions in the subsection are not entirely new and a number of dedicated 
model-independent analysis exist in the literature~\cite{Cirelli:2008pk,Bi:2009uj,Feng:2013zca,Calore:2013yia,Yuan:2013eja,Cholis:2013psa,Jin:2013nta,Kajiyama:2013dba,Yuan:2013eba,Yin:2013vaa,Dienes:2013lxa,Ibarra:2013zia}. Here we consider on the 
$X_{\mu}\rightarrow l^+l^-,\;l=(e,\mu,\tau)$, and shall give a detailed explanation on 
why each channel can or cannot fit the data in a qualitative manner. 
We shall give simple illustrations without doing a precise global fit to the data, 
focusing only on the $E_k>10$GeV range.

Since we assume $X_{\mu}$ can decay to leptons only, it will give rise to indirect
signatures in cosmic $e^{\pm}$, which can be conveniently discussed in terms of two 
observables: the total flux $\Phi_{e^{-}+e^{+}}$ and the positron fraction 
$\Phi_{e^{+}}/\Phi_{e^{-}+e^{+}}$.
Each $\Phi$ is the sum of background flux and the contribution from
dark matter decay. The $e^{\pm}$ background fluxes of interstellar
origin can be  parametrized analytically as \cite{fermilat,Ibarra:2009dr}
\begin{eqnarray*}
\Phi_{e^{-}}^{\mathrm{bkg}}\left(E\right) & = & \left(\frac{82.0\times E^{-0.28}}{1+0.224\times E^{2.93}}\right)\mathrm{GeV^{-1}m^{-2}s^{-1}sr^{-1},}\\
\Phi_{e^{+}}^{\mathrm{bkg}}\left(E\right) & = & \left(\frac{38.4\times E^{-4.78}}{1+0.0002\times E^{5.63}}+24.0\times E^{-3.41}\right)\mathrm{GeV^{-1}m^{-2}s^{-1}sr^{-1},}
\end{eqnarray*}
where $E$ is in GeV unit. For the flux from VDM decay, we
calculate it with modifying $\texttt{micrOMEGAs}$\cite{micromegas}. The production
rate is given by 
\begin{equation}
Q\left(E,\vec{r}\right)=\frac{\rho\left(\vec{r}\right)}{M_{\mathrm{DM}}\tau_{\mathrm{DM}}}\frac{dN^{e^{\pm}}}{dE},\label{eq:source2}
\end{equation}
$dN^{e^{\pm}}/dE$ is the energy spectrum function, $M_{\mathrm{DM}}=M_{X}$ in our
discussion, $\tau_{\mathrm{DM}}$ is the lifetime of $X_{\mu}$ and
$\rho\left(r\right)$ is the density profile of dark matter.  We use the NFW profile for
the decaying DM, too.

\begin{figure}
\includegraphics[width=0.48\textwidth]{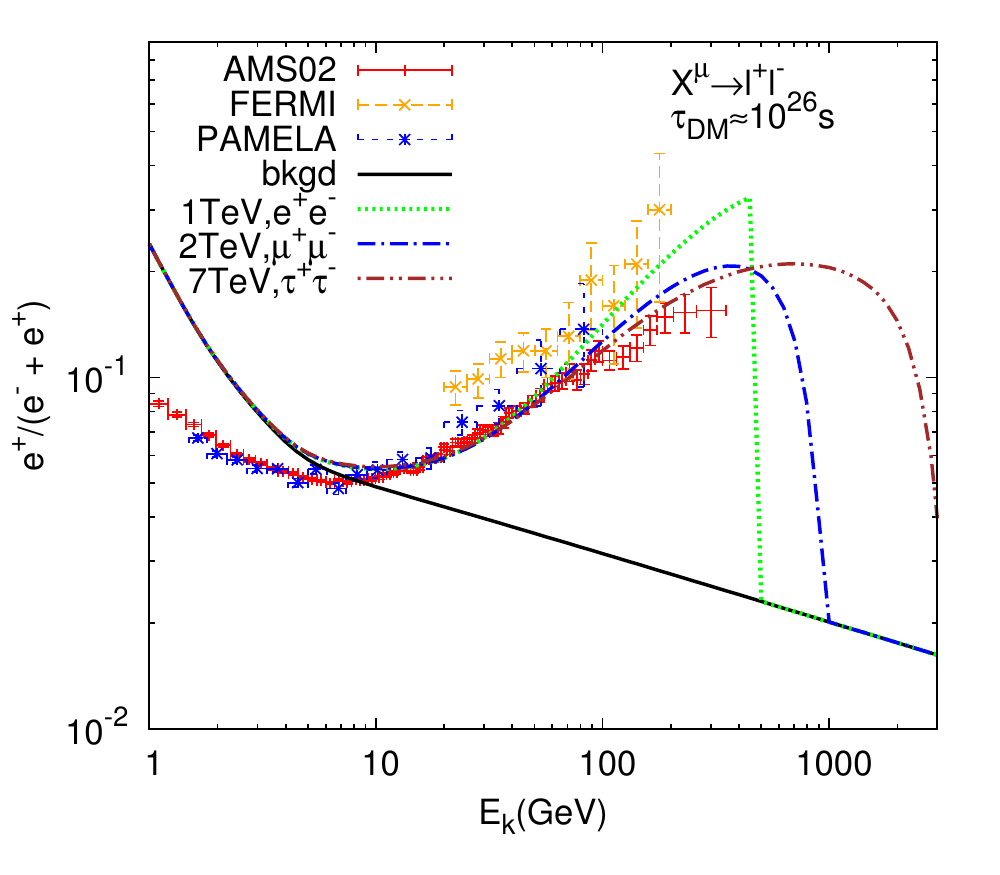}
\includegraphics[width=0.48\textwidth]{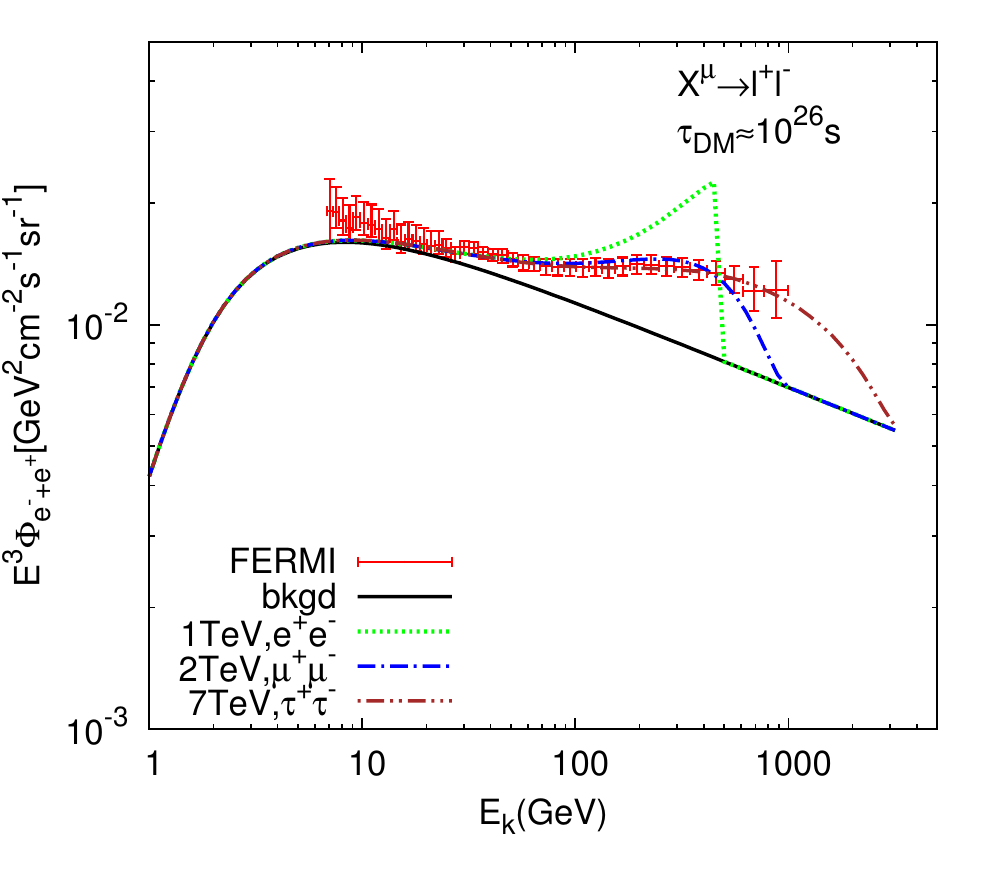}
\caption{These figures show the spectra of the $e^{+}$ fraction
and $e^{\pm}$ total flux from the decay, $X^{\mu}\rightarrow l^{+}l^{-},\; l=e,\;\mu,\;\tau$ 
and lifetime $\tau_{\textrm{DM}}$ is chosen to $(4,2,0.7)\times10^{26}$s, respectively. 
These parameters are chosen for the illustration purpose only. 
See details in text.\label{fig:cosmicray}}
\end{figure}

In Fig.~\ref{fig:cosmicray}, we show  the spectra of the positron fraction and 
$\Phi_{e^{-}+e^{+}}$ for individual decay channel, 
$X^{\mu}\rightarrow l^{+}l^{-},\; l=e,\;\mu,\;\tau.$
To compare with experimental observation, we have also shown the data from 
PAMELA,  Fermi and AMS02. In the low energy range $E_k <10$  GeV, it is known
that solar wind can have significant effects on the charged particles,
the so-called solar modulation which depends strongly on the solar
activity. Since the uncertainty for the background flux in this range  is large, 
we shall not discuss the spectra for $E_k <10$ GeV any further in this paper. 
For $E_k >10$ GeV the mass of $X_{\mu}$ and lifetime  $\tau_{\mathrm{DM}}$ are 
chosen to give relatively better fit with the data . 

From green dotted curves in Fig.~\ref{fig:cosmicray}, we can see that
$X^{\mu}\rightarrow e^{+}e^{-}$ can not be consistent with the positron
fraction and the total flux simultaneously. The reason is that the spectrum
of $e^{\pm}$ from $X_{\mu}$'s decay is very hard and too sharp around
$E=M_{X}/2$, and it is inconsistent with the Fermi data on the total flux.   

In case of $X_\mu \rightarrow \mu^{+}\mu^{-}$, the situation is much better 
as shown in  the blue dot-dashed curves for $M_{X}=2$ TeV and 
$\tau_{\mathrm{DM}}\simeq2\times10^{26}\mathrm{s}$.
Since the produced $\mu^{\pm}$ undergoes subsequently three-body decay 
$\mu^{\pm}\rightarrow e^{\pm}+\nu_{e}+\nu_{\mu},$  the resulting $e^{\pm}$ spectrum 
from the VDM decay becomes much softer compared with the 
$X^{\mu}\rightarrow e^{+}e^{-}$ case.

In the $\tau^{+}\tau^{-}$ case, the $e^{\pm}$ spectrum is even softer
than the $\mu^{+}\mu^{-}$,  since only one third of $\tau^{\pm}$ decay
to $\mu^{\pm}$ and $e^{\pm}$. Other $\tau$'s decay hadronically
into lighter mesons which then decay further to pions, followed by 
$\pi^{\pm}\rightarrow\mu^{\pm}+\nu_{\mu}$
and $\pi^{0}\rightarrow2\gamma$. However, the spectrum's softness
could be compensated with an even heavier $X_{\mu}$. As we show in
the brown double-dot-dashed lines of Figure. \ref{fig:cosmicray}, $M_{X}=7$ TeV and 
$\tau_{\mathrm{DM}}\simeq0.7\times10^{26}\mathrm{s}$ can give
a good fit with the data. We shall note $m_X = 7$ TeV lies in the  boundary of 
previous constraints as the perturbativity limits $g_{X}\lesssim1.5$  which further 
set the upper bound $M_{X}\lesssim7$ TeV to give correct relic density. 

{\it $\gamma$-ray Constraints}:  It is well known that the cosmic $\gamma$-ray is 
an important constraint on both pair-annihilating and decaying dark matter.
If combined with gamma-ray constraint for decaying dark matter based on Fermi-LAT  
data \cite{Cirelli:2008pk,Cirelli:2009dv,Ackermann:2012qk,Ackermann:2012rg,Cirelli:2012ut,
Gomez-Vargas:2013bea}, the only viable channel is $X_{\mu}\rightarrow\mu^{+}\mu^{-}$. 
The reason is that for the $e^{+}e^{-}$ channel all the lost energy goes to photons.  
For the $X_\mu \rightarrow \tau^{+}\tau^{-}$ case, the decay products has a lot of 
$\pi^0$ which then all decay to 2$\gamma$, while for $X_\mu \rightarrow \mu^{+}\mu^{-}$ 
a large part of muon energy is carried by the neutrinos in the decay product of muon, 
leaving less energy for electron to radiate $\gamma$. Interpretation and constraints 
after AMS02 have been discussed in \cite{Feng:2013zca,Calore:2013yia,Pearce:2013ola,DeSimone:2013fia,Yuan:2013eja,Ibe:2013nka,Cholis:2013psa,Jin:2013nta,Kajiyama:2013dba,Yuan:2013eba,Masina:2013yea,Yin:2013vaa,Feng:2013vva,Liu:2013vha,Dienes:2013lxa,Bergstrom:2013jra,Ibarra:2013zia,Geng:2013nda}, which would not change the $\gamma$-ray constraints.

\subsection{Implications for thermal VDM with mass $\sim 2$ TeV}

Accounting for the positron excess observed by PAMELA and AMS02 through  
thermal VDM ($\sim 2$ TeV) decaying into $\mu^+ \mu^-$  
will restrict the parameter space of the renormalizable Lagrangian, which is one of 
the main results of this paper.
As we have shown in sec.~\ref{sec:tevDM}, for $\mathcal{O}$(TeV) VDM,  
the thermal relic density can pin down the gauge coupling $g_X \simeq 0.76$ in the dark sector (see Fig.~\ref{fig:constraints} and Eq.~(\ref{eq:gxmx}). Then only $M_{H_2}$ 
and the exact mixing angle $\alpha$ are not fixed,  but they are correlated with 
and constrained by Higgs data, DM direct searches, BBN and thermalization assumption, displayed in Fig.~\ref{fig:xenon}. Taking $M_{H_2}\simeq 300\mathrm{GeV}$ as an example, 
we have $\sin \alpha \lesssim 0.3 \;\textrm{ and }\; \lambda_H\gtrsim 0.129.$

With all the current constraints taken into account and taking $M_{H_2}\geq 150$GeV, 
sizable deviation is possible for the Higgs self-coupling as shown in Fig.~\ref{fig:hself}.  Precise measurements of Higgs self couplings at the future colliders then could fix 
$M_{H_2}$ and $\alpha$. Then we can  predict the $X_\mu$-nucleon cross section for DM 
direct searches and our model gets testable. If we further require that the electroweak 
vacuum is stable up to the scale $\Lambda \sim 10^{15} - 10^{16}$ GeV, 
then all parameter space can be probed by XENON1T.  This is an interesting and 
important result within our approach on decaying VDM thermalized by Higgs portal 
interaction.

\section{Summary}
In this paper, we have investigated the phenomenology (mainly focusing on indirect 
signatures) of a vector dark matter $X_{\mu}$ in the framework of Higgs portal model, 
enlarging the SM gauge group  $G_{SM}$ by a dark $U(1)_{X}$. 
We first discussed the primary cosmic rays, including  $\gamma$-ray and neutrino fluxes, 
from $X_{\mu}$-$X_{\mu}$ annihilation and compare the spectra in several cases.  
In order to explain the  positron excess observed by PAMELA and AMS02, 
we then focus on the TeV scale $M_{X}$ and show it can evade all the constraints 
from the Higgs data, relic density, perturbativity and dark matter direct search.   
Signals from heavy $X_\mu$ pair annihilation into leptons are well below the background 
and data. Since having the boost factor from the Sommerfeld enhancement is strongly 
constrained and basically ruled out by CMB, we then turn to the signatures from 
$X_{\mu}$'s decay for explanation of the positron excess observed by PAMELA and AMS02. 

We have also presented all the independent dim-6 operators that involve both 
standard model and dark sector  particles, and that are invariant under the  
$G_{SM}\times U(1)_{X}$  gauge symmetry.   After the breaking of 
$G_{SM}\times U(1)_{X}$,   the VDM $X_{\mu}$  can decay to the SM particles.
A TeV VDM $X_{\mu}$ can also explain the excess of positron fraction recently observed 
in PAMELA, FERMI and AMS02 experiments. We give an example model to implement
a leptophilic interaction and show the relevant indirect signature.
It is shown that $X_{\mu}\rightarrow e^{+}e^{-}$ gives a spectrum too hard to
explain the observation while $X_{\mu}\rightarrow\mu^{+}\mu^{-},\;\tau^{+}\tau^{-}$
can be consistent with both positron fraction and the total $e^{\pm}$  flux. However, 
if we take the constraints from the gamma ray, then only 
$X_{\mu}\rightarrow \mu^{+}\mu^{-}$ is viable.

Our study presented in this paper is different from other model independent analysis 
of cosmic rays in the literature. We demonstrated explicitly that thermalization 
of the VDM is possible for $\sim$ TeV scale VDM,  and then considered the VDM decays 
into a lepton pair.  The indirect searches for cosmic rays can determine the VDM mass, 
which then fixes the $U(1)_X$ gauge coupling for giving the thermal relic density. 
The only left two correlated parameters are the mass of second scalar and its mixing 
angle with Higgs. These two can be further probed by future collider searches, for instance, 
precision measurement of Higgs self coupling or production of the second scalar, and DM direct searches at XENON1T for example. 
The physical observables we have discussed systematically 
in this paper are complementary to each other and testable in terrestrial experiments.   
Similar analyses could be done for other types of decaying DM assuming they are  
thermalized through some interactions (such as Higgs portal or singlet portal interactions
~\cite{Baek:2013dwa}).

\begin{acknowledgments}

This work is supported in part by National Research Foundation of Korea (NRF) Research Grant 2012R1A2A1A01006053 (SB,PK,WP,YT), and by the NRF grant funded by the Korea government (MSIP) 
(No. 2009-0083526) through  Korea Neutrino Research Center at Seoul National University (PK).

\end{acknowledgments}

\end{document}